\documentclass[]{spieman}  

\usepackage{amsmath,amsfonts,amssymb}
\usepackage{graphicx}
\usepackage{aas_macros}
\usepackage{setspace}
\usepackage{tocloft}
\usepackage[normalem]{ulem}

\title{Automated data processing architecture for the Gemini Planet Imager Exoplanet Survey}

\author[a,*]{Jason J. Wang}
\author[b]{Marshall D. Perrin}
\author[c]{Dmitry Savransky}
\author[d]{Pauline Arriaga}
\author[e,f,g]{Jeffrey K. Chilcote}
\author[a]{Robert J. De Rosa}
\author[h,i]{Maxwell A. Millar-Blanchaer}
\author[j,k]{Christian Marois}
\author[l]{Julien Rameau}
\author[m,b]{Schuyler G. Wolff}
\author[c]{Jacob Shapiro}
\author[e]{Jean-Baptiste Ruffio}
\author[n]{J\'{e}r\^{o}me Maire}
\author[o]{Franck Marchis}
\author[a]{James R. Graham}
\author[e]{Bruce Macintosh}
\author[p]{S. Mark Ammons}
\author[e]{Vanessa P. Bailey}
\author[q]{Travis S. Barman}
\author[r]{Sebastian Bruzzone}
\author[s]{Joanna Bulger}
\author[t]{Tara Cotten}
\author[l]{Ren{\'e} Doyon}
\author[a,u]{Gaspard Duch\^{e}ne}
\author[d]{Michael P. Fitzgerald}
\author[v]{Katherine B. Follette}
\author[w]{Stephen Goodsell}
\author[x]{Alexandra Z. Greenbaum}
\author[y]{Pascale Hibon}
\author[z]{Li-Wei Hung}
\author[aa]{Patrick Ingraham}
\author[a]{Paul Kalas}
\author[n]{Quinn M. Konopacky}
\author[d]{James E. Larkin}
\author[ab]{Mark S. Marley}
\author[r,ac]{Stanimir Metchev}
\author[o,e]{Eric L. Nielsen}
\author[ad]{Rebecca Oppenheimer}
\author[p]{David W. Palmer}
\author[ae]{Jennifer Patience}
\author[p]{Lisa A. Poyneer}
\author[b]{Laurent Pueyo}
\author[b]{Abhijith Rajan}
\author[w]{Fredrik T. Rantakyr\"o}
\author[ae]{Adam C. Schneider}
\author[b]{Anand Sivaramakrishnan}
\author[t]{Inseok Song}
\author[b]{Remi Soummer}
\author[aa]{Sandrine Thomas}
\author[h]{J. Kent Wallace}
\author[ae]{Kimberly Ward-Duong}
\author[af]{Sloane J. Wiktorowicz}

\affil[a]{Department of Astronomy, UC Berkeley, Berkeley CA, 94720, USA}
\affil[b]{Space Telescope Science Institute, 3700 San Martin Drive, Baltimore MD 21218 USA}
\affil[c]{Sibley School of Mechanical and Aerospace Engineering, Cornell University, Ithaca, NY 14853, USA}
\affil[d]{Department of Physics and Astronomy, UCLA, Los Angeles, CA 90095, USA}

\affil[e]{Kavli Institute for Particle Astrophysics and Cosmology, Stanford University, Stanford, CA 94305, USA}
\affil[f]{Dunlap Institute for Astronomy \& Astrophysics, University of Toronto, 50 St. George St, Toronto ON M5S 3H4, Canada}
\affil[g]{Department of Physics, University of Notre Dame, 225 Nieuwland Science Hall, Notre Dame, IN, 46556, USA
}

\affil[h]{Jet Propulsion Laboratory, California Institute of Technology, 4800 Oak Grove Dr., Pasadena CA 91109, USA}
\affil[i]{NASA Hubble Fellow}
\affil[j]{National Research Council of Canada Herzberg, 5071 West Saanich Road, Victoria, BC V9E 2E7, Canada}
\affil[k]{University of Victoria, 3800 Finnerty Rd, Victoria, BC, V8P 5C2, Canada}
\affil[l]{Institut de Recherche sur les Exoplan\`{e}tes, D\'{e}partment de Physique, Universit\'{e} de Montr\'{e}al, Montr\'{e}al QC H3C 3J7, Canada}
\affil[m]{Physics and Astronomy Department, Johns Hopkins University, Baltimore MD, 21218, USA}

\affil[n]{Center for Astrophysics and Space Science, University of California San Diego, La Jolla, CA 92093, USA}
\affil[o]{SETI Institute, Carl Sagan Center, 189 Bernardo Ave.,  Mountain View CA 94043, US}

\affil[p]{Lawrence Livermore National Laboratory, 7000 East Ave., Livermore, CA 94550, USA}
\affil[q]{Lunar and Planetary Lab, University of Arizona, Tucson, AZ 85721, USA}
\affil[r]{Department of Physics and Astronomy, Centre for Planetary Science and Exploration, The University of Western Ontario, London, ON N6A 3K7, Canada}
\affil[s]{Subaru Telescope, NAOJ, 650 North A’ohoku Place, Hilo, HI 96720, USA}
\affil[t]{Department of Physics and Astronomy, University of Georgia, Athens, GA 30602, USA}
\affil[u]{Univ. Grenoble Alpes/CNRS, IPAG, F-38000 Grenoble, France}
\affil[v]{Amherst College Department of Physics and Astronomy, Merrill Science Center, 15 Mead Drive, Amherst, MA 01002, USA}
\affil[w]{Gemini Observatory, 670 N. A'ohoku Place, Hilo, HI 96720, USA}
\affil[x]{Department of Astronomy, University of Michigan, Ann Arbor, MI 48109, USA}
\affil[y]{European Southern Observatory , Alonso de Cordova 3107, Vitacura, Santiago, Chile}
\affil[z]{Natural Sounds and Night Skies Division, National Park Service, Fort Collins, CO 80525, USA}
\affil[aa]{Large Synoptic Survey Telescope, 950N Cherry Ave., Tucson, AZ 85719, USA}

\affil[ab]{Space Science Division, NASA Ames Research Center, Mail Stop 245-3, Moffett Field CA 94035, USA}

\affil[ac]{Department of Physics and Astronomy, Stony Brook University, Stony Brook, NY 11794-3800, USA}
\affil[ad]{American Museum of Natural History, Depratment of Astrophysics, Central Park West at 79th Street, New York, NY 10024, USA}
\affil[ae]{School of Earth and Space Exploration, Arizona State University, PO Box 871404, Tempe, AZ 85287, USA}
\affil[af]{The Aerospace Corporation, 2310 E. El Segundo Blvd., El Segundo, CA 90245}

\cftpagenumbersoff{figure}
\cftpagenumbersoff{table} 
\begin{document} 
\maketitle

\begin{abstract}
The Gemini Planet Imager Exoplanet Survey (GPIES) is a multi-year direct imaging survey of 600 stars to discover and characterize young Jovian exoplanets and their environments. We have developed an automated data architecture to process and index all data related to the survey uniformly. An automated and flexible data processing framework, which we term the Data Cruncher, combines multiple data reduction pipelines together to process all spectroscopic, polarimetric, and calibration data taken with GPIES. With no human intervention, fully reduced and calibrated data products are available less than an hour after the data are taken to expedite follow-up on potential objects of interest. The Data Cruncher can run on a supercomputer to reprocess all GPIES data in a single day as improvements are made to our data reduction pipelines. A backend MySQL database indexes all files, which are synced to the cloud, and a front-end web server allows for easy browsing of all files associated with GPIES. To help observers, quicklook displays show reduced data as they are processed in real-time, and chatbots on Slack post observing information as well as reduced data products. Together, the GPIES automated data processing architecture reduces our workload, provides real-time data reduction, optimizes our observing strategy, and maintains a homogeneously reduced dataset to study planet occurrence and instrument performance.
\end{abstract}


{\noindent \footnotesize\textbf{*}Jason Wang,  \linkable{j-wang@berkeley.edu} }

\begin{spacing}{1}   

\section{INTRODUCTION}
\label{sec:intro}  
The Gemini Planet Imager (GPI) is a high contrast imaging instrument on the Gemini South telescope designed to directly image young, recently formed exoplanets and their planet-forming environments\cite{2007arXiv0704.1454G, Macintosh2014}. To suppress the glare of the bright host star to see faint planets and circumstellar material, GPI is equipped with a high-order adaptive optics (AO) system to correct for atmospheric turbulence\cite{Poyneer2016} and an apodized-pupil Lyot coronagraph to suppress diffraction from the star\cite{Soummer2011}. To image and characterize planets and disks, an infrared integral field spectrograph (IFS) sits behind the AO system and coronagraph. The IFS uses a microlens array to disperse light from a 2.8~$\times$~2.8 arcsecond field of view onto a HAWAII 2-RG detector in one of five filters, $Y$, $J$, $H$, $K1$, and $K2$, where $K$-band has been split into two filters. The IFS supports both a spectral mode, with a spectral resolving power $R$ between 30 and 80 depending on the wavelength, and a polarimetry mode that allows for broadband imaging polarimetry using a Wollaston prism and a rotating waveplate\cite{Chilcote2012, Larkin2014, Perrin2015}. 

The GPI Exoplanet Survey (GPIES) is a multi-year survey of 600 young, nearby stars with GPI to discover new planets, characterize the orbits and atmospheres of directly-imaged planets, constrain giant planet occurrence rates, and resolve planet-forming environments. GPIES is searching for planets in spectral mode in $H$-band around 600 stars, following-up and characterizing planet candidates in multiple wavelength bands at multiple epochs, and looking for resolved circumstellar material in polarized, scattered light around a subset of stars.

From both a technical and scientific perspective, GPIES requires an automated infrastructure to handle the data associated with the survey. Technically, both the size and complexity of data processing requirements make manual bookkeeping impractical. Over several years, GPIES will accumulate more than 30,000 raw frames of science data and a roughly equal number of raw frames of calibration data. While this is not close to the scale of the largest surveys in astronomy, it is large enough that we need automated methods to track and organize the data. Additionally, data reduction for GPIES is non-trivial, requiring a pipeline to reduce IFS data and complex algorithms to model and subtract out the diffracted light from the host stars. For each raw IFS frame, we must calibrate and extract $\sim$35,000 microspectra from the detector to make a single spectral datacube. If we consider each wavelength slice of each spectral datacube as an individual frame for data processing, which effectively is how it gets treated, then we will need to subtract the diffracted light of the star from over one million frames using multiple algorithms to mitigate algorithmic biases. To maintain data consistency, we need an automated system to reprocess all of our data as upgrades and bug fixes are applied to our data reduction pipelines.

Scientifically, we need the infrastructure to make sense of all the data from GPIES. Consistently processed data is important to maintain a homogeneous dataset to use for statistical planet occurrence studies. Being able to easily access specific data from across the entire survey is important to understand trends in instrument performance, inform observing strategies, and prioritize future instrument upgrades. Finally, having real-time processing during the observing night is extremely advantageous for optimizing observing strategy. Fully-reduced data that is processed and displayed in real time helps observers assess the data quality being achieved. It also allows observers to identify candidate companions within an hour of the observations so that the most promising candidates can be followed up the same night or the following night. 

\begin{figure}
    \begin{center}
    \includegraphics[width=16cm]{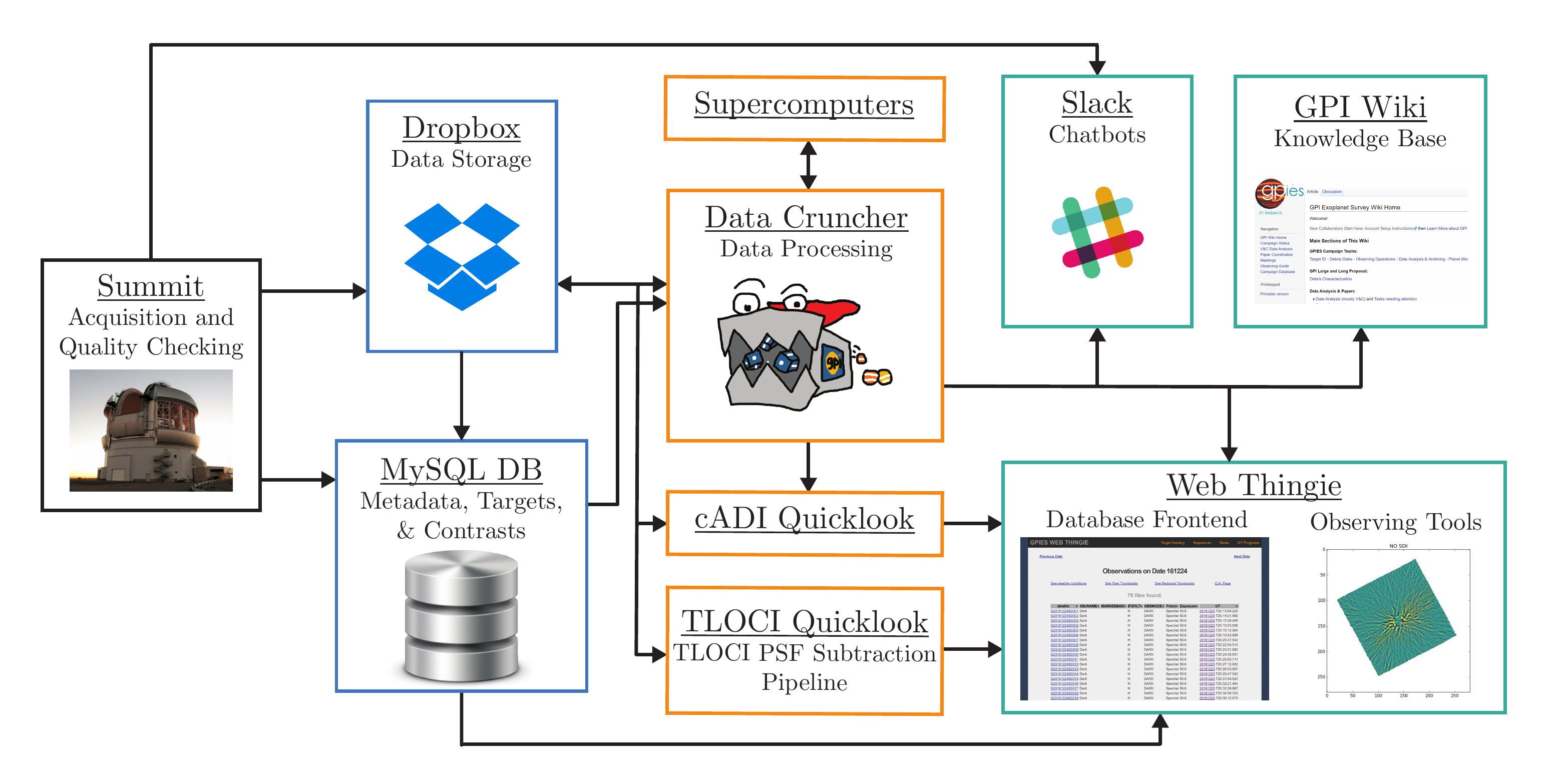}
    \end{center}
    \caption
    { \label{fig:data-infrastructure} 
    Schematic of the GPIES automated data processing infrastructure. Boxes represent the different components of the infrastructure that are described in Section \ref{sec:data-infrastructure}. The boxes are colored so that black represents the telescope, blue represents data storage, orange represents data processing modules, and teal represents user-facing services. Arrows indicate the flow of data or information from one component to another.}
\end{figure}

Thus, we have built an automated data processing infrastructure capable of storing all of the data, processing all science and calibration data in real-time, reprocessing the entire campaign as necessary, and providing powerful but easy-to-use tools to search across all data from the survey,. The individual components and how they fit together are displayed in Figure \ref{fig:data-infrastructure}. 

The different components that comprise the data infrastructure, including the Data Cruncher, a key component that automatically reduces all GPIES data, are described in Section \ref{sec:data-infrastructure}. In Section \ref{sec:utility}, we describe the benefits of this automated architecture for observing, data processing efficiency, instrument performance, and survey statistics. Finally, in Section \ref{sec:future-steps}, we conclude with plans for future steps to take for the automated data processing architecture. We also include two appendices: Appendix \ref{sec:data-reduction} details all of the data reduction steps for GPIES, which we have automated, and Appendix \ref{sec:code-data-cruncher} describes the software implementation of the Data Cruncher.

\section{The Components of the Automated GPIES Data Infrastructure}
\label{sec:data-infrastructure}
Briefly, we will summarize the data reduction detailed in Appendix \ref{sec:data-infrastructure} to give context for this data infrastructure. For each star observed in GPIES, the GPI Data Reduction Pipeline (DRP) is used to turn raw IFS data into spectral datacubes with two spatial dimensions and one spectral dimension. To see planets in the spectral datacubes, ``stellar PSF subtraction" algorithms take advantage of the angular differential imaging (ADI)\cite{Marois2006_adi} and spectral differential imaging (SDI)\cite{Marois2000} observing techniques to remove the point spread function (PSF) of the star, which for these coronagraphic images is the diffracted light of the star behind the occulting mask. Two stellar PSF subtraction algorithms are used: \texttt{pyKLIP} and cADI, which are described in Appendix \ref{sec:spec-psfsub}. Then, the sensitivity of each dataset to planets is computed. In this paper, we will use the term ``contrast" to refer to the flux ratio between the faintest planet we can detect and its star, and the term ``contrast curve" to refer to our achieved contrasts in a dataset as a function of projected separation from the star. Stars with debris disks are imaged in broadband polarimetry mode, and the data are processed to remove the unpolarized stellar light to look for polarized light from small dust grains. In order to process the science data, calibration files for both our spectral and polarimetry mode data are also reduced.

To automatically download, index, process, and display all this data from GPIES, multiple components have been integrated together to form the automated data infrastructure. Figure \ref{fig:data-infrastructure} provides a schematic of the various parts of the infrastructure, which we will discuss in detail here. Roughly, the infrastructure is divided into data acquisition and storage, data processing, and front-end interfaces. In Section \ref{sec:data-acq-and-store}, we will describe the data acquisition and storage: data tools for quicklook quality checking on the summit (Section \ref{sec:summit}), the database that stores all of the metadata, target information, and planet sensitivity curves (Section \ref{sec:database}), and Dropbox, which stores the raw and reduced images as well as AO telemetry (Section \ref{sec:dropbox}). Next in Section \ref{sec:data-cruncher}, we will describe the functionality of the Data Cruncher, the automated data processing framework that automatically processes all science and calibration data from GPIES. Quicklook software, including one from the separate TLOCI pipeline\cite{Marois2014}, that creates real time displays of the current observing sequence are discussed in Section \ref{sec:quicklook}. The rest of this section is dedicated to our front-end interfaces. Our main user-facing front-end is the Web Thingie, which hosts views into our database as well as observing tools (Section \ref{sec:webthingie}). We also describe how collaborative tools such as Slack and our internal wiki are integrated into our automated data infrastructure (Section \ref{sec:collab-tools}).

\subsection{Data Acquisition and Storage}
\label{sec:data-acq-and-store}
A substantial amount of data is being generated by GPIES. About three years into the survey, we have accumulated $\sim$23,000 raw science files and $\sim$26,000 raw calibration files. Including reprocessed data, we have generated $\sim$27,000 quicklook datacubes, $\sim$5,600 reduced calibration files, $\sim$80,000 science-grade datacubes, $\sim$68,000 stellar-PSF-subtracted images, and $\sim$61 million contrast curve data points. In addition to the science data, we also have $\sim$73,000 raw telemetry files produced by the AO system to monitor its performance, the observing status of all 600 targets in the survey, and also information on the target stars themselves. To handle all of the data while also making it available to the entire collaboration, we use a combination of Dropbox and a MySQL database to store the data.

\subsubsection{Summit Quicklooks and Data Download}
\label{sec:summit}
After new IFS data are taken on the summit, an instance of the GPI DRP running on the summit uses the GPI DRP Autoreducer module to automatically perform quicklook reductions of the data. These quicklook reductions allow observers to assess data quality and, for science data, provide a contrast curve to demonstrate the sensitivity achieved in a single frame of data for understanding observing conditions. If the data are rendered unusable due to issues like AO control loops opening or wind-shake moving the star off of the coronagraph mask, the observer can log that particular file as bad through the GPItv interface in the GPI DRP. The summit quicklook reductions provide observers the basic tools to assess data quality so that observing can continue in the unlikely case the observatory network connection fails and renders the rest of the data infrastructure ineffective. During an observing sequence, the observer also periodically takes AO telemetry data every 5-10 minutes to allow for further analysis of AO system performance. 

To move the data off the summit, automated scripts download the raw and quicklook science data as well as the AO telemetry data to a server located at Cornell University that hosts the MySQL database. While AO telemetry data is downloaded during daytime hours to avoid saturating the network bandwidth during the night, IFS frames are written at a rate of fewer than one per minute, so it is downloaded in real time along with the bad files log without using significant network resources. The server at Cornell University then uploads the data and metadata to Dropbox and the database respectively.

\subsubsection{The Database}
\label{sec:database}
After receiving new data, the Cornell server adds entries for all of the data into a MySQL database. For all of the science and calibration data taken by the IFS, header information and metadata gets uploaded into the raw and reduced data tables. The raw table contains one row for each raw file, with one column for each of the fields in the file headers (e.g., observing mode, wavelength band, time of observation), along with a column for the unique identifier (ID) for each file. The reduced table contains one row for each reduced file, produced either by the quicklook GPI DRP on the summit or by the Data Cruncher afterwards. The reduced data table contains information after some data processing has happened such as whether the data product is a quicklook or science-grade reduction, the sensitivity achieved at some fiducial separations, flux calibration conversions, as well as a unique ID for the reduced file. To link the reduced data to their original raw data products, a third ``Raw2Reduced" table is a two column table where each row associates one raw file ID with one reduced file ID. Multiple rows can specify that multiple raw files went into producing one reduced file. Similarly, multiple rows could contain multiple reduced files that all used a single raw file, but not necessarily exclusively that raw file. Contrast curves associated with a given reduction (either quicklook contrast curves from the summit or Data Cruncher contrast curves) are stored in a contrast curve table where each row is the contrast for a given reduced file at a given separation. Lastly, the bad files log is appended into the observing notes table, which contains notes on each data frame. A row consists of a text comment, whether the bad file flag was enabled, the user that submitted it, and time it was submitted, and the associated raw data file it pertains to. Multiple rows can correspond to a single file to give a history of bad-file marking.

Similarly, the AO data has a raw data table and a reduced data table in a similar fashion as the IFS data, but with AO-relevant metadata stored in each of the tables (e.g., the wavefront sensor frame rate, spatial filter size, and seeing). To make the most out of the AO data, there exist tables that link the raw AO data to reduced AO data and the raw AO telemetry to raw IFS data. The latter table allows AO performance metrics to be linked to final sensitivity to planets.

In the database there also exist tables for GPIES targets. To keep track of the large GPIES target list, we use several target tables. The basic target table contains information on the target such as the celestial coordinates, an estimate for the age of the system, and the target ID number. Then, there is a table with all of the photometry in the literature on each target, a table of name aliases for each target, a table of binary companions associated with each target, a table with spectral energy distribution (SED) fits for each target, a table with the Gemini Observing Tool observing sequence numbers for each target, and a table of observing statuses (e.g., observed, incomplete, candidate companion). 

\subsubsection{Dropbox}
\label{sec:dropbox}
After the data are logged into the database, the raw and quicklook reductions are uploaded onto Dropbox. The raw and summit-reduced quicklook data are put into a read-only directory, as they are not meant to be modified again. After the raw data are synced to Dropbox, the Data Cruncher can then read and process the data automatically. The automatically reduced data from the Data Cruncher also gets synced back onto Dropbox, and then their metadata also stored into the database. All of the AO telemetry is also stored and synced on Dropbox in its own folder.

Sometimes, it is necessary to reduce data by hand, especially when there is an astrophysical source that requires further analysis. In these cases, individuals also upload their own reductions and analysis outputs onto Dropbox so that they can be shared with others. The only restriction is that these data products lie outside of the automatically generated directories, to avoid being erased or moved. This is because after the Data Cruncher reprocesses the entire campaign, all of the automatically generated folders gets moved into an archive section of Dropbox, and new reductions are moved into its place. The archive contains subdirectories, each with an older version of all of the data reduced in the survey by the Data Cruncher.

\subsection{The Data Cruncher}
\label{sec:data-cruncher}
\begin{figure}
    \begin{center}
    \includegraphics[width=16.5cm]{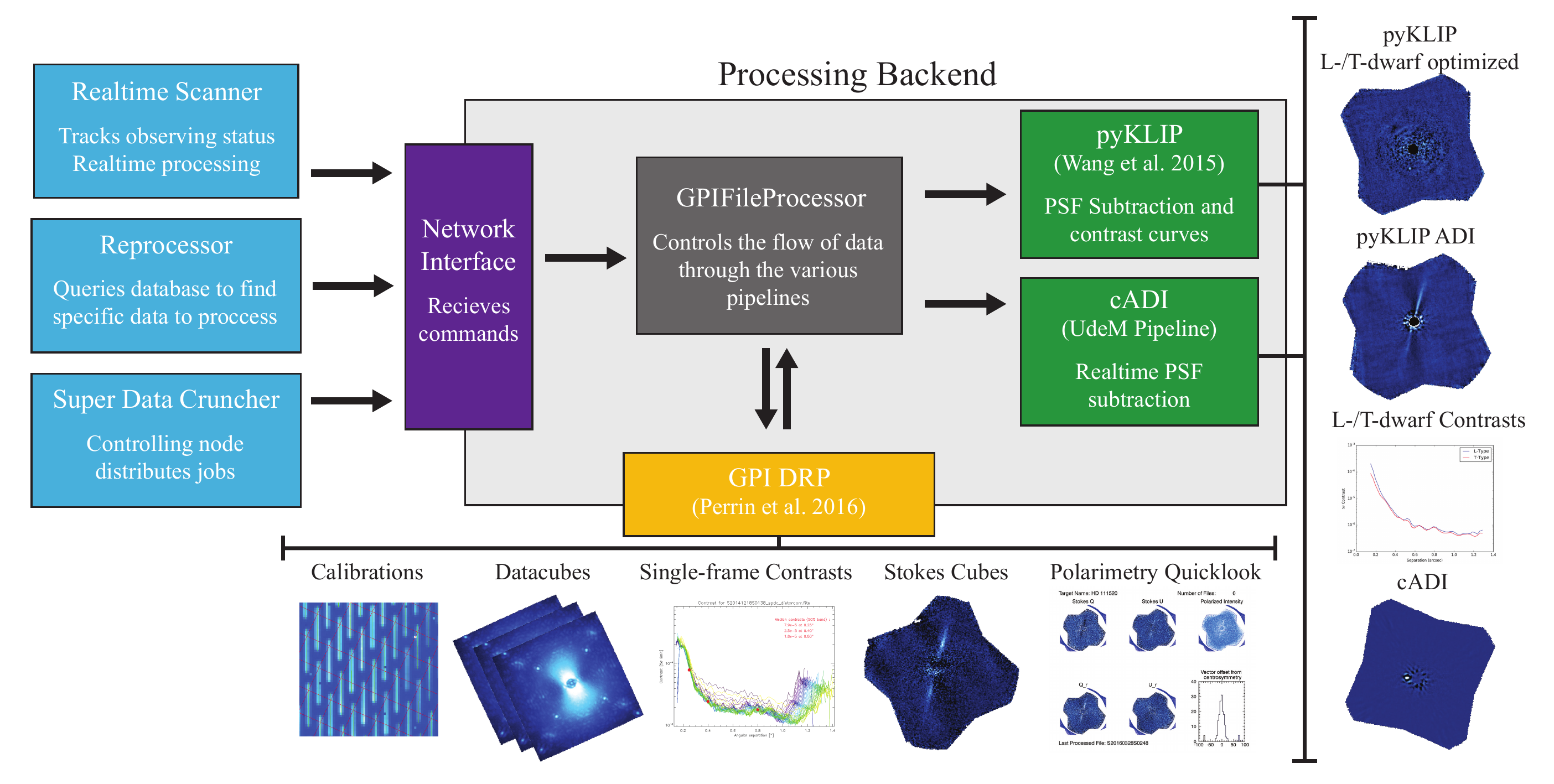}
    \end{center}
    \caption
    { \label{fig:data-cruncher-schematic} 
    Schematic of the Data Cruncher architecture. Boxes represent the different modules of the Data Cruncher, which are discussed in Appendix \ref{sec:code-data-cruncher}. The light blue boxes represent the various Instructors that are described in Appendix \ref{sec:instructors}. The large light gray box represents the Processing Backend, which consists of several smaller modules. Arrows indicate the flow of data or information from one module to another. Samples of the different data products are shown and grouped by the pipeline that produced them (either the GPI DRP or the stellar PSF subtraction pipelines). }
\end{figure}

The Data Cruncher is a Python framework that automates all of the data processing steps detailed in Appendix \ref{sec:data-reduction}. The Data Cruncher is comprised of two parts: the ``Processing Backend" (Appendix \ref{sec:processing-backend}) runs the various data reduction pipelines and controls the flow of data through the pipelines, while a series of ``Instructors" (Appendix \ref{sec:instructors}) sends commands to the Processing Backend about what data to process. Figure \ref{fig:data-cruncher-schematic} illustrates how the Instructors and Processing Backend fit together to generate data products for the survey. By separating the code between a unit that focuses on the data processing and a unit that focuses on what data to process, we are able to develop a modular and flexible framework. This allows us to scale the Data Cruncher arbitrarily: multiple Instructors can use the same Processing Backend to process data for multiple purposes, or a single Instructor can talk to many Processing Backends when a large amount of data needs to be processed in a parallel fashion.

The detailed software implementation of the Data Cruncher, including the details on the Instructors and Processing Backends, is discussed in Appendix \ref{sec:code-data-cruncher}. At a high abstraction level, one can think of the Data Cruncher as being able to produce all of the desired data products given some raw GPI data. Here, we will focus on the functionality of the Data Cruncher for GPIES. 

\subsubsection{Real-Time Reductions}
\label{sec:real-time-processing}
When a raw data file is uploaded to Dropbox, the ``Realtime Scanner" Instructor is alerted and decides how to instruct the Processing Backend to process the file depending on the context in which the file is taken. For example, if a new raw science frame is downloaded, the Realtime Scanner will just instruct the generation of a datacube from the raw 2-D file and the updating of the quicklook stellar PSF subtractions (described in Section \ref{sec:quicklook}) if the object name in its file header is the same as the previous file's. However, it will also send instructions for the full stellar PSF subtraction of the current data sequence if the object name is different, since this indicates that the observers have moved on to a new target. The Realtime Scanner is able to handle all standard observing procedures for all spectral, polarimetry, and calibration data taken as part of GPIES. The only exception is that the Realtime Scanner is not programmed to handle processing thermal background frames and subtracting them from the entire sequence. However, we find that thermal background subtraction is only necessary in $K$-band for extended sources such as circumstellar disks, for which we need to distinguish between the large-scale astrophysical emission of the disk from the smooth thermal background. $K$-band imaging of disks is not a science-goal of GPIES, so this only happens when the Data Cruncher processes queue programs led by GPIES members. In these cases, manual reductions need to be performed. 

On a 32-core machine with AMD Opteron 6378 processors clocked at 2.3~GHz, the Data Cruncher generates quicklook reductions within 1 minute of receiving the data, generates the first \texttt{pyKLIP} reduction 10 minutes after a sequence finishes, and generates fully-calibrated contrast curves for both L- and T-type planets 1 hour after a sequence is complete. 

\subsubsection{Reprocessing Individual Datasets}
Sometimes, a single dataset needs to be reprocessed in order to exclude bad frames or to fix bugs in the reduction pipelines. The Reprocessor module allows us to query the database for certain datasets, find the corresponding files, and process them. For a specified target, the user can request for data taken at a given wavelength band, on a given date, in a given observing mode, and in a given observing program (whether GPIES or an affiliated proposal by GPIES team members) to be processed. The user can also specify whether to process the data completely from raw data, or to use the reduced datacubes and only run specified post-processing algorithms. For example, a user can ask to reprocess all data on a target taken in polarimetry mode when upgrades to the GPI DRP's polarimetry reduction functions happen. A user could also ask to regenerate the contrast curves for a given target at a given band to exclude files where the data quality was poor. 

\subsubsection{Reprocessing the Entire Campaign}
When the Data Cruncher runs on a supercomputer, we call the framework the Super Data Cruncher. Many nodes, each a single computer, are requested, and each node runs one instance of the Processing Backend to parallelize the embarrassingly parallel problem of reprocessing the entire campaign. A single node is designated the controlling node and sends commands to all of the Processing Backends to split up the work. The Super Data Cruncher has successfully run on the Edison and Cori machines operated by the National Energy Research Scientific Computing Center (NERSC) as well as the Comet machine operated by the San Diego Supercomputer Center (SDSC) as part of the Extreme Science and Engineering Discovery Environment (XSEDE)\cite{xsede}. On Comet, using 30 nodes, with 24 cores per node for a total of 720 cores, we used 16,560 CPU hours, corresponding to 22.9 hours of wall-clock time, to reprocess all survey data up to the end of 2016 (17,008 frames of raw IFS data).

\subsection{Quicklooks}
\label{sec:quicklook}
During the observing night as data are taken and reduced in real time, quicklook tools translate the data into easy-to-view images that are constantly refreshed as new data comes in. Currently, we run three quicklook tools, which update every minute as new data are taken. Each quicklook tool syncs images to a certain location on Dropbox so that it can be picked up and displayed on web pages on the Web Thingie.

\begin{figure}
    \begin{center}
    \includegraphics[width=16.5cm]{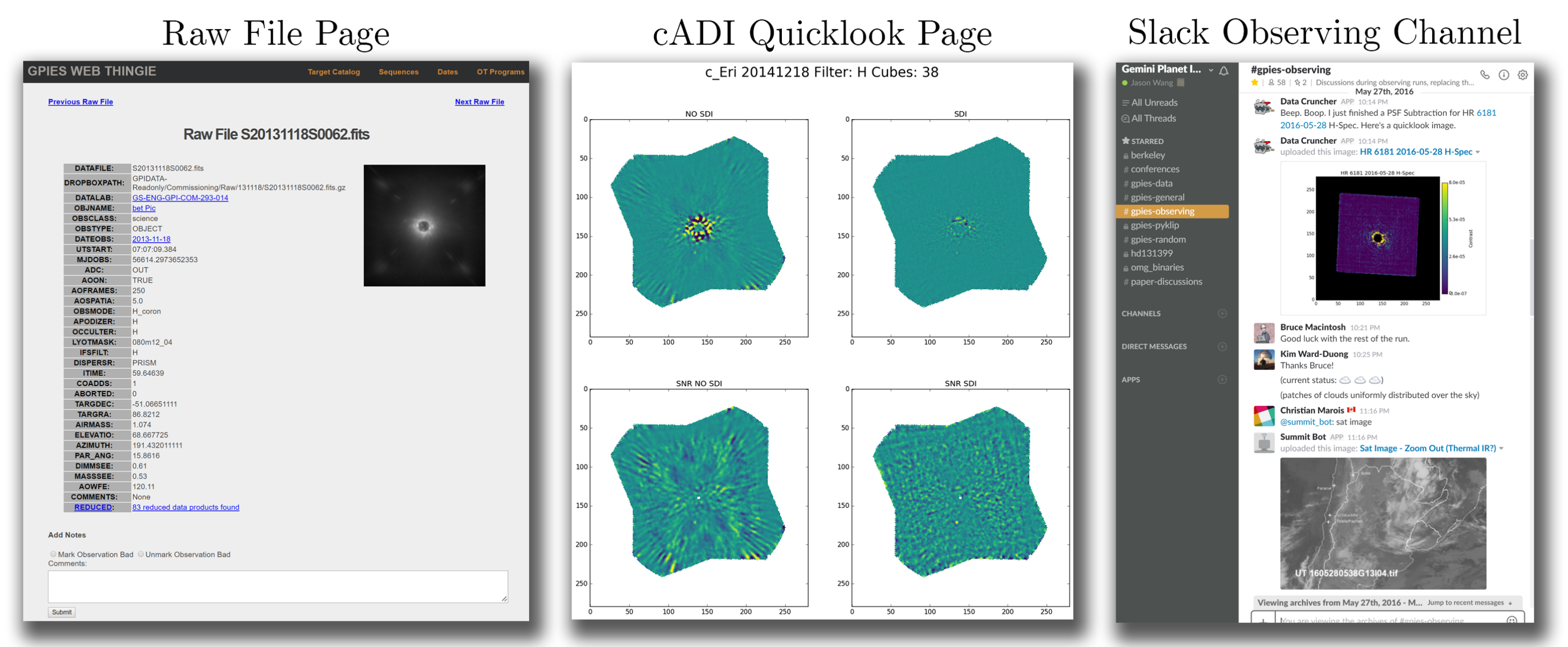}
    \end{center}
    \caption
    { \label{fig:frontend} 
    Example screenshots of the user-facing web front ends that are integrated with the data processing infrastructure. {\it Left}: A dynamically generated page from the Web Thingie for one single raw data file. This is one of multiple types of pages the Web Thingie can generate. {\it Center}: The real-time cADI quicklook display that is automatically generated and updated in real time on the Web Thingie. {\it Right}: An example of the observing support provided by the Slack chatbots. Even though the weather was poor, the Data Cruncher continued to produce science-grade stellar-PSF-subtracted data and the Summit bot provided updates on the weather conditions as requested.  }
\end{figure}

The first is the cADI quicklook tool. The Data Cruncher produces the two cADI reductions of spectral mode data described in Appendix \ref{sec:spec-psfsub} every time new frames are taken and uploaded to Dropbox. The cADI quicklook tool takes these reductions and creates a four-panel plot to display stellar-PSF-subtracted images (Figure \ref{fig:frontend}). The top two images of the plot are just the data without any further processing. The bottom two images are the signal-to-noise ratio (SNR) maps of the data after convolving the data with a Gaussian function and weighing each pixel by the inverse of the noise in pixels of similar separations from the star. 

The second is the TLOCI quicklook display, which uses stellar-PSF-subtracted spectral mode data from the TLOCI pipeline\cite{Marois2014}, independent of the Data Cruncher. The TLOCI pipeline typically uses the quicklook spectral datacubes made on the summit so that it does not rely on the Data Cruncher for datacubes, but can be modified to point at the spectral datacubes made by the Data Cruncher. The TLOCI quicklook display offers four plots, two for L-dwarf reductions and two for T-dwarf reductions. For each spectral type, there is a reduction done by a quick simple code, and a slower, but more sophisticated code. TLOCI quicklook display also offers a simple planet detection code and quicklook astrometry and spectra of candidates flagged by the planet detection code. 

The last is the polarimetry quicklook display generated by the Data Cruncher through the GPI DRP. This updates after every four images (corresponding to a full cycle of waveplate positions) and displays images from both the linear and radial Stokes cubes produced by the Data Cruncher that are described in Appendix \ref{sec:pol-psfsub}. 

\subsection{The Web Thingie: Web Front End for Database and Observing Tools}
\label{sec:webthingie}
Between all of the raw, reduced, calibration, and AO telemetry data generated by the GPIES campaign, it quickly becomes hard to track down the specific files taken on specific dates, and the associated reductions. Instead of requiring users to make SQL queries to the database themselves when they need to look for data, we have created a user-friendly web front end, named the Web Thingie, to allow users to find data, mark files as bad, pick targets, and look at real-time stellar PSF subtractions. 

The Web Thingie is run on an Apache-based web server gateway interface using Flask, a Python, Werkzeug-based web development framework. It uses Jinja2 to dynamically generate web pages from database queries. For example, instead of static web pages, the pages that list all raw files taken on each date are generated from one template page with the date populated by the user requested date and rows of a raw files table populated from a database query. The Web Thingie can a page for any individual raw or reduced file stored on Dropbox, for raw files taken on a specific date, for raw files corresponding to a specific object, for raw files of a given calibration file type, for a calendar indicating dates under each month in which GPI data was taken, among others. 

Typically users are interested in data taken on a specific date or of a specific object, so going to those pages will list all data from that date or object including the instrument configuration, wavelength, time, and the data quality of the file. Once they do click on an individual raw file, a web page following the same template as the one shown in Figure \ref{fig:frontend} is generated. There is a thumbnail image of the raw data frame, a list of important header keywords, a link to a page that lists all the reduced data products that used this raw data frame, existing notes on this data frame, and the ability to add more notes and mark whether this frame is bad or not. If the data frame is bad, a large text banner in red font on the top of the page indicates that. When the user clicks through to the list of all the reduced data, they can scroll through and click on the desired reduced file to go to its page for more information on the reduction. 

Each target also gets its own page to help observers during a run. On the top of the page, there are links to the SIMBAD page for the target, the internal wiki page for the target, and the finder chart for crowded field stars. Then, the status of the target is listed regarding whether spectral or polarimetry data needs to be taken. Basic information on the star as well as binary companions and their separation and position angle are listed next, to help observers pick the right star at the telescope. The page ends with a series of data quality plots, which are generated dynamically. The first set of plots shows single-frame contrast, AO wavefront error, and satellite spot flux as a function of image number for all observing sequences on this target. These metrics indicate how conditions change over an observing sequence and allow for comparison to previous observing sequences on the same target. The second set of plots displays the histograms of all single-frame contrasts for all targets in the database, all targets within 0.25 magnitudes, and all previous data from the same star. It also displays the 16th, 50th, and 84th percentile single frame contrasts for all targets within 0.25 magnitudes. This allows observers to quickly compare the quality of the data from the summit quicklook reductions that they are currently taking and see how they compare to the rest of the campaign, allowing for a straightforward assessment of data quality. 

To maintain target selection flexibility, dynamically generated target selection tools are also available on the Web Thingie. At any given time of the night, a dynamic target list page automatically orders targets by a parameterized score\cite{mcbride2011experimental, savransky2013campaign} that is a combination of the inherent proprieties of the system (e.g., age, distance) as well as how easily observable the target is (e.g., targets that transit directly overhead need to be observed just before transit for maximum field rotation for ADI). The list can also be trimmed to only look at bright stars, or to avoid targets that require the telescope to point in the current direction of the wind. The list also flags whether polarimetry observations are needed, alerts observers to binary stars with colorfully flashing star icons to ensure that the correct star is observed, and compares how much field rotation is achieved observing it immediately versus the maximum achievable field rotation in a one-hour observing sequence. Additionally, any target can be plotted on the ``Nightly Planning Tool" page, which plots the elevation, cumulative field rotation, and instantaneous field rotation over time to determine the feasibility of observing the target at any given time. An offline target list also exists in case the Web Thingie is ever inaccessible.

Lastly, the Web Thingie hosts three quicklook stellar PSF subtraction monitors to allow observers to search for candidates and assess stellar-PSF-subtracted data quality in real time as discussed in Section \ref{sec:quicklook}.

\subsection{Collaborative Tools}
\label{sec:collab-tools}
To streamline collaboration across an international team, a series of tools are used. To improve their utility, we have integrated some of these tools with the automated infrastructure. 

\subsubsection{Wiki}
\label{sec:wiki}
An internal wiki run on the Dokuwiki platform is used for documentation. The wiki is used as a knowledge-base for information like troubleshooting the instrument. For data analysis, a target page is automatically generated for each target, with an observing summary of that target automatically generated by the Data Cruncher. Users can then post additional data analysis numbers, such as astrometry and photometry of candidates, as well as summaries of the current state of data analysis.

\subsubsection{Slack}
\label{sec:slack}
The messaging application Slack is used to communicate efficiently between collaborators located all across the world. The ``\texttt{\#gpies-observing}" channel allows for focused discussion during the current observing run. Slack also has a programmatic API, allowing for the development of software integrations. As shown in Figure \ref{fig:frontend}, The Data Cruncher has a chatbot interface (source code available at \linkable{https://github.com/semaphoreP/datacruncher\_slackbot}) which posts \texttt{pyKLIP} reductions as it finishes them to \texttt{\#gpies-observing}. When messaged, it also can retrieve images of previous \texttt{pyKLIP} reductions on GPIES targets, tell the time in any time zone, provide sunrise and sunset times for Gemini South, and display the current moon phase using an emoji. For telescope and observing information, a ``Summit Bot," also shown in Figure \ref{fig:frontend}, was forked from the Data Cruncher chatbot to run on the Gemini South summit computers and thus can access the state of the telescope, weather monitors, and cloud cameras, which have proved extremely useful for remote observing when no one from the GPIES team is on the summit. Together, the two chatbots assist in taking observations and analyzing data, reducing the manpower needs for executing the survey. They also allow GPIES members outside of the observing team to easily stay updated on observing status: within half an hour after an observing sequence on a target is complete, GPIES members can look at the fully reduced data the Data Cruncher chatbot posted onto Slack, without anyone in GPIES having to do any data processing. 

\section{Utility of the Data Infrastructure}
\label{sec:utility}
A fully automated data architecture from start to finish brings several benefits to GPIES regarding observing efficiency, data processing, data archiving, instrumental performance, and survey statistics. 

\subsection{Tools for Observers}
A whole suite of tools has been developed to inform observers in real time, reducing the complexity of decisions that need to be made during the night. Before and during the night, the Summit Bot on Slack can be messaged to display environmental monitors and webcams from the summit. This allows the observers to look at the current time series of data from seeing monitors as well as look at in which directions in the sky there might be clouds. Since all of this data appear on the \texttt{\#gpies-observing} Slack channel, members of GPIES outside from the core observing team can chime in, providing advice and improving the cohesion of the team. Once conditions are good for observing, the automated target selection tool on the Web Thingie allows for flexible scheduling of targets, giving observers the highest priority target at any given time, after accounting for pointing restrictions.  
Once a target is picked, and data are being acquired, the GPI DRP on the summit generates quicklook datacubes to check for star alignment and image quality. Obviously bad data, such as when the AO control loops open, can be marked as bad in the GPI DRP and the bad file flags will be propagated into the database. Quicklook single-frame contrasts for the new data can be compared to the histogram of all single-frame contrast of similar targets. This allows observers to understand in what percentile the current observing conditions are, since single-frame contrasts are the best predictors of final stellar-PSF-subtracted contrast for a whole dataset\cite{Bailey2016}. 

As we accumulate frames on a target, the quicklook stellar PSF subtraction pages on the Web Thingie update in real-time. This is particularly useful when following up known systems or candidate planets. Because of different observing conditions affecting AO performance and stellar PSF stability, the exposure times to achieve a desired SNR for confirmation or characterization of astrophysical sources will vary. The quicklook reductions, even though they do not achieve the best stellar PSF subtraction, give us an excellent measure. If a candidate companion does not show up in the quicklook tools after a standard one-hour follow-up sequence, observers can decide to add an extra hour to the follow-up observations. Contrarily, if conditions are excellent and the quicklooks already detect the source at high SNR, then observers can choose to cut a long sequence short, and move on to other targets. 

After a sequence is complete, observers can use the Web Thingie to change the target status flag, flagging it as complete, incomplete, or requiring follow-up of candidates. Typically, ten minutes after a sequence is complete and with no human intervention, the Data Cruncher Slack bot posts the final \texttt{pyKLIP} stellar-PSF-subtracted data onto \texttt{\#gpies-observing}, allowing for convenient viewing of the previous dataset and discussion of possible candidates in the data.

\subsection{Data Processing Speed and Consistency}
To make inferences from a large survey, it is important that the data are processed uniformly with the latest calibrations and bug-fixes. With such a large amount of data over several years, an automated architecture to handle all of the various data types and data processing needs can mitigate errors and reduce the workload, particularly for junior members of the GPI collaboration who often are the ones reducing the data. 

Using the Data Parser functionality of the GPI DRP, the Data Cruncher is able to identify and process all calibration files taken before or during an observing run. Without the need to trigger human intervention to process these data, we always have the latest calibration files processed before each run and available on Dropbox. This also means that there is no need to reprocess the data again immediately after the run. The real-time reductions generated by the Data Cruncher are of science-grade and are used in papers. 

Because of this, we have an extremely quick turnaround when we do detect interesting objects. For example, when GPIES discovered HR 2562 B\cite{Konopacky2016}, we could watch in real time using the quicklooks on the Web Thingie as the companion appeared in our data over the course of an hour. 
Preliminary analysis of the companion indicated a peaky $H$-band spectrum indicative of a bound, L-type companion rather a background star, which likely has a monotonically smooth spectrum. We had confidence in our preliminary analysis since the wavelength calibration and spectral datacubes produced by the Data Cruncher were of science grade. We knew we wanted to follow it up immediately to characterize the spectrum at other wavelength bands, and we did so at the next possible opportunity during the same observing run. Similar situations have happened with background objects, for which spectra and astrometry were extracted within hours of observation and candidates were identified as background objects within the same night by comparing with archival data. The fast turnaround time on our data processing means we can quickly identify and prioritize the most interesting objects found by GPIES. This becomes important when planets are rare and weather conditions are bad. 

When upgrades or bug-fixes to the various pipelines occur, the Super Data Cruncher is able to reprocess the entire campaign on a supercomputer within a single day. This way, we can maintain a homogeneous reduction process for all data, and observations collected at different times are not subject to varying systematic errors associated with changes in the reduction process. Eliminating these systematic bugs in a uniform manner is necessary when performing statistical analysis on the campaign.

\subsection{Large Scale Data Analysis}
A large dataset that is uniformly processed enables a wide range of analyses that allow us to better understand planets, our algorithms, and GPI itself. A critical final goal of the survey is to place limits on the occurrence of giant planets at Solar System scales by comparing the number of planets imaged to the sensitivity to planets based off contrast curves generated by the Data Cruncher. Because the contrast curves are also stored in the database, both single-frame and final stellar-PSF-subtracted contrasts can be correlated with target and atmospheric keywords stored in the data headers. This allows us to understand instrument performance by linking metrics generated by the AO system to metrics expressing planet sensitivity generated by the Data Cruncher \cite{Poyneer2016, Bailey2016}. For example in \citenum{Bailey2016}, we were able to construct the histogram of GPI performance for stars of different brightness as well as show GPI's AO correction is dependent on the speed of atmospheric turbulence rather than just seeing. As another example, the measurements of the apparent stellar polarization obtained throughout GPIES can be used to improve the characterization of the instrumental polarization over previous methods \cite{Millar-Blanchaer2017}. 

The Data Cruncher cannot satisfy every data analysis need, since there are always additional analyses to be done. While rarely are datacubes regenerated, manual stellar-PSF-subtractions are done for scientifically interesting objects to optimize the reduction for that object. Additionally, as new planet detection algorithms are developed, such as Forward Model Matched Filter (FMMF)\cite{Ruffio2017}, they are run using the uniformly processed datacubes made by the Data Cruncher to characterize algorithm performance. While the Data Cruncher is not the solution to all data reduction needs, it offers a solid infrastructure for others to run analysis on. 

Also, because all data from GPIES are processed uniformly and available on Dropbox in an organized fashion, publicly releasing GPIES data after the campaign will be straightforward as sharing the link to the Dropbox folder containing the most up-to-date reductions from the Data Cruncher.

Most importantly, all of this data analysis is possible without a large effort by many members of GPIES reducing the data by hand. A standard 1~hour GPIES observing sequence takes 1~hour to manually generate all of the reduced data products. If follow-up observations are included, $\sim$430 datasets have been observed, so assuming an eight-hour workday, it would have required $\sim$54 days worth of work to manually reduce all of the data taken so far once. Reprocessing the campaign would be incredibly time-consuming since each reprocessing would take just as many work hours. Instead, the processing and reprocessing of GPIES data is accomplished with minimal time investment by humans. This has allowed GPIES members to focus on higher-level analysis and writing papers rather than processing data.

\section{Future Steps}
\label{sec:future-steps}
The successful design and implementation of the automated data architecture has been one of the major accomplishments of the GPIES collaboration. A few additional features remain desirable. Our top priority is to transition from planet detections by-eye to the FMMF algorithm, which can automatically flag candidate companions with low false positive rates\cite{Ruffio2017}. Automated application of FMMF has not been implemented, partially due to the substantial computation cost of the current version of the algorithm. Simpler planet detection algorithms have been implemented and tested in the Data Cruncher and TLOCI pipeline, but these create more work than they save because they flag too many false positives. After an automated planet detection algorithm is implemented into the Data Cruncher, automated astrometry and spectrophotometry of candidates will follow. However, given the rate of planet candidate detections (a few per observing run), it has not been an efficient use of time to implement all of these features.

A natural application of the GPIES architecture would be to other direct imaging surveys, past, present, and future. Given the similar data products and data processing needs, it would be relatively straightforward to adapt many of our tools for these surveys. Using such an infrastructure to uniformly process all the data would ensure archival data is processed with the latest stellar PSF subtraction algorithms and contrast curves are removed of biases between data reduction pipelines when combining results from multiple surveys.

Although GPIES has very specific data processing and technical needs, some of the infrastructure from GPIES could be broadly applicable to other surveys of similar scales, where the data volume is large enough that doing everything by hand is impractical, but small enough that it lacks a dedicated team of professionals to manage all of the data. Pieces of the infrastructure or general principles could be applicable to other surveys. Much of the architecture described here can credit its success to the integration and design of the entire system. Features such as linking all reduced data products to raw data products to instrumental and weather condition keywords in a database can useful for other surveys, but perhaps in different forms (e.g., there might be several intermediate data products that need to be tracked). Another area to emphasize are the real-time processing and observing tools that reduce the complexity of observing and streamline optimization of observing strategy. Aggregating observing statistics and providing real-time observing conditions in convenient ways such as a website or Slack can benefit other surveys too.

In the mean time, the GPIES team will enjoy the automation of the data processing for the rest of the survey. The lack of updates to the Data Cruncher recently ($\sim$2 commits a month on average to our Git repository in 2017) is evidence that the infrastructure is running smoothly, and that the team is reaping the benefits of the work put into the automation. 

\appendix    

\section {Data Reduction for GPIES}
\label{sec:data-reduction}

\subsection{Calibration Data}
\label{sec:calib}
All processing of calibration files uses the Data Parser module of the open-source and publicly available GPI DRP \cite{Perrin2014, Perrin2016}. The Data Parser can be used through its GUI interface or programmatically in IDL using the \texttt{parse\_fileset\_to\_recipes()} function that is associated with the \texttt{parsercore} object. Either way, the Data Parser is automatically able to identify and create data reduction steps for all of our calibration data. 

In addition to the standard observatory calibration sequence of dark frames that are taken at a variety of exposure times, GPIES takes a sequence of 80 60~s dark frames at the end of each night, as most of the campaign uses 60~s exposure times. Each sequence of darks of the same exposure time is combined to produce one dark frame\cite{Ingraham2014}.

Wavelength calibration frames, or ``wavecals," are taken at two times: during the day and right before each science sequence. During the day, deep sequences are taken with both the argon and xenon arc lamps at each band as part of the standard observatory calibration sequence. These are processed to be ``master wavecals" since they are of high SNR and allow for the wavelength solution of each microspectra to be computed individually. Before each science sequence, GPIES takes a 30~s argon arc lamp observation at the band the science data is to be observed in (except in $K$-band, where an $H$-band arc lamp frame is taken). This frame is processed using the ``Quick Wavelength Solution" recipe template, which measures shifts in the microspectra due to instrument flexure by using a master wavecal and computing the global offset in both X and Y necessary to shift the master wavecal to align with the microspectra of the argon arc lamp. This frame is only used to correct for flexure, and relies on the master wavecal for the rest of the wavelength solution. The accuracy of the GPI wavelength solution is $< 2$~nm in each band \cite{Wolff2014}.

For polarimetry data, we need calibration files that specify the locations and the shapes (fit as 2-D Gaussians) of the polarization spot pairs for each lenslet. Each spot in the pair corresponds to the intensity in a given polarization channel of a spatial pixel in a polarimetry datacube. The calibration files provide the required mapping between each spot on the detector and its location within the final datacube (where the 3rd dimension represents the polarization state). Calibration files are generated in each band for the polarimetry spot calibration using the flat field lamp of the Gemini calibration unit, GCAL, and processed using the GPI DRP Data Parser. 

\subsection{Spectral Mode}
The majority of data from GPIES comes in the form of IFS data, where the raw data consists of $\sim$35,000 microspectra spread across the detector.

\subsubsection{Constructing Spectral Datacubes}
\label{sec:spec-datacubes}
To extract the microspectra into a more useful data product, we use the GPI DRP to construct spectral datacubes. The datacubes are constructed and calibrated using the steps listed in Table \ref{tab:spdc}, which are slightly different than those of the default recipes offered as part of the GPI DRP. Each step corresponds to a primitive, one specific reduction task, that can be run in the GPI DRP.

While most of the primitives are straightforward, there are a few points to clarify. The wavelength calibration that is loaded in is typically the argon arc snapshot taken right before each science sequence described in Section \ref{sec:calib}. In $K$-band, the master wavecal is loaded in since $H$-band arc snapshots are taken instead. To then correct for instrument flexure, we use the ``BandShift" method of the ``Update Spot Shifts for Flexure" primitive. For sequences not in $K$-band, this only corrects for shifts of the microspectra due to flexure caused by  changes in the elevation of the telescope between the snapshot arc and the current science frame. For a $K$-band sequence, this feature also corrects for the shift between the current data and the master $K$-band wavecal that is loaded in by measuring the offset between the $H$-band master wavecal and the $H$-band arc snapshot that was taken before the sequence. This requires the $H$- and $K$-band master wavecals to be taken together without the telescope having been moved in between so that they experience the same flexure. 

Due to a few large clusters of bad pixels on the detector, we are not able to interpolate over all bad pixels in the 2-D frame. Some bad pixels are propagated to the datacube, where we can use neighboring spatial pixels in the datacube that are far apart in the 2-D frame to fix the remaining bad pixels. 

The satellite spots, four fiducial spots created by diffraction of starlight off of a grid and centered on the location of the star \cite{Marois2006_satspots,Sivaramakrishnan2006}, are used for locating the occulted star and calibrating the data photometrically. The GPI DRP measures the position and flux of each satellite spot, and stores them to the header of the datacube. The mean of the positions of the four spots in each wavelength channel is also written into the header as the location of the star at each wavelength. It is up to the various post-processing pipelines whether to use these numbers or to recalculate them. In particular, it is important to measure the flux of the satellite spots and astrophysical sources in the same way to mitigate biases, so often the satellite spot fluxes are recomputed. However, the values stored here are used to calibrate the single-frame contrast curves, so they are important to log. 

\begin{table}
\centering
\caption{GPI DRP processing steps to make a spectral datacube.}
\begin{tabular}{| l | p{10cm} | }
 \hline
 \textbf{Primitive Name} & \textbf{Purpose}  \\
 \hline 
 Load Wavelength Calibration & Reads in the appropriate wavelength calibration file \\
 \hline
 Subtract Dark Background & Finds an appropriate dark frame and subtracts it from the raw 2-D image \\
 \hline
 Update Spot Shifts for Flexure & Corrects for shifts in the microspectra due to instrument flexure\cite{Wolff2014} \\
 \hline
 Interpolate Bad Pixels in 2D Frame & Identifies and fixes bad pixels\cite{Ingraham2014} \\
 \hline
 Assemble Spectral Datacube & Extracts the data into a 3-D cube using a 3-pixel wide moving box \\
 \hline
 Interpolate Wavelength Axis & Interpolates the wavelength dimension to be 37 equally-spaced channels \\
 \hline
 Interpolate Bad Pixels in Cube & Fixes any remaining bad pixels using spatially nearby pixels \\
 \hline
 Correct Distortion & Corrects for optical distortion. The datacubes are saved after this step, and following steps only modify the header information. \\
 \hline
 Measure Satellite Spot Locations & Automatically finds satellite spots using a computer vision algorithm and fits their locations with a Gaussian\cite{savransky2013, Wang2014} \\
 \hline
 Filter Datacube Spatially & Subtracts off a 15-pixel median filter from the image to high-pass filter the image for the following steps. The filtered image is not saved to disk.  \\
 \hline
 Measure Satellite Spot Peak Fluxes & Uses a Gaussian matched filter and the location of the satellite spots in the header to measure the flux of each satellite spot \\
 \hline
 Measure Contrast & Measures the single-frame contrast. Then it saves the 1-D contrast curve to a FITS file and contrasts at three fiducial separations to the header of the datacube \\
 \hline
\end{tabular}
\label{tab:spdc}
\end{table}

\subsubsection{Stellar PSF Subtraction}
\label{sec:spec-psfsub}
Taking all of the datacubes obtained in a sequence on a given target, we use ADI and SDI to distinguish the light of the star from the light of other astrophysical sources, so that we can model and subtract out the stellar point spread function. To avoid dependence on a single stellar PSF subtraction algorithm, we use both the classical ADI algorithm (cADI)\cite{Marois2006_adi} and \texttt{pyKLIP}\cite{Wang2015}, an open-source Python implementation of Karhunen-Lo\`{e}ve Image Projection (KLIP)\cite{Soummer2012,Pueyo2015}, to provide two separate subtractions of the stellar PSF. The cADI pipeline is quick so it is used for real-time reductions, while the \texttt{pyKLIP} pipeline is used for our final sensitivity analysis. 

Using cADI, we perform two subtractions. The first is a basic ADI reduction where the images are collapsed into broadband images using the mean, high-pass filtered using a Fourier filter with a smooth cutoff frequency of four spatial cycles, subtracted with the mean of the broadband image sequence, rotated North-up, and collapsed in time to form one 2-D stellar-PSF-subtracted image.
The second reduction is similar to the first except before collapsing into broadband images, it includes an SDI pre-processing step where in each spectral cube, each frame is subtracted by the median of all frames in the same spectral cube. The latter subtraction is better suited for planets with sharp molecular absorption features in the spectrum, while the former works better for all other cases. 

Using \texttt{pyKLIP}, we perform three reductions to search for astrophysical signals: one general reduction, one optimized for T-dwarfs with strong molecular absorption features, and one optimized for circumstellar disks. In the first general-purpose reduction, we first high-pass filter the images using an 11-pixel full-width half-maximum (FWHM) Gaussian filter in Fourier space to remove broad features caused by the seeing and wind-shake while minimizing attenuation of high-frequency point source signals of planets. Gaussians with smaller FWHMs were found to subtract out too much point source signal. Then, the images are duplicated once for each spectral channel to be aligned to a common center and magnified using bicubic spline interpolation so that the speckles are aligned at that wavelength (i.e., so that the data at that wavelength do not need to be magnified through interpolation). Then, we break the images into nine annuli that increase logarithmically with separation and break each annulus into four azimuthal sectors. For each frame in a given sector, we build our Karhunen-Lo\`{e}ve (KL) modes using the 300 most correlated reference frames where a hypothetical planet at the center of the sector would have moved at least one pixel due to a combination of ADI field rotation and SDI rescaling of the speckles. One pixel, which is about a third of the FWHM of the planet PSF in $H$-band, was chosen empirically to maximize the SNR of potential planets. Since the number of KL-modes to use varies depending on the dataset, we save images that use 1, 10, 20, 30, and 50 KL-modes to reconstruct and subtract off the stellar PSF. The images are all rotated to be North-up. We then save six separate data products. Five are spectral datacubes, one for each KL-mode cutoff, where we have collapsed all the data in time. The last one is a KL-mode cube where we have collapsed the images in both time and wavelength, leaving the third dimension to be the number of KL-modes used to model the stellar PSF. This cube allow for quick visual inspection of the dataset to determine what the optimal number of KL-modes to use is and to search for planets.

The second reduction optimized for T-dwarfs is similar to the first except for two changes. First, the exclusion criterion of one pixel is weighted so that reference images where the planet is faint due to molecular absorption is preferred. This means that even if the hypothetical planet did not move one pixel due to ADI and SDI, as long as it is faint enough at that wavelength compared to at the wavelength we are trying to perform stellar PSF subtraction, the image will be included in the reference library. Appendix A.1.3 of~\citenum{Ruffio2017} explains this in quantitative detail and illustrates this selection process for a typical GPI dataset. Second, the KL-mode cube is collapsed in wavelength using a weighted mean where the weights are the fluxes for each spectral channel of a model T-dwarf atmosphere. 

The third reduction is optimized for disks and again is similar to the first reduction except for three changes: the images are not high-pass filtered to avoid removing disk structure, the images only use ADI so only images at the same wavelength are used to build the reference data, and more conservative KL-mode cutoffs of 1, 3, 10, 20, 50 are saved. 

\subsubsection{Contrast Curves}
\label{sec:spec-contrastcurves}
After stellar PSF subtraction, we compute the sensitivity to planets in the dataset. For GPIES, we typically are sensitive to planets with L-type and T-type spectra. Due to the strong molecular absorption of methane in T-type spectra, we have improved sensitivity to T-type planets since we can use all the spectral channels where the planet is dim due to methane absorption to be more aggressive in modelling and subtracting out the stellar PSF without significant self-subtraction of the planet itself (see Section \ref{sec:spec-psfsub}). Thus, due to the differing sensitivities, we produce contrast curves for both L- and T-type planets. 

For L-type and T-type planets, we use the general reduction and the T-type optimized reduction respectively to estimate the level of the noise in the image. We then cross correlate each reduction with a Gaussian function that has a FWHM of 3.5 pixels to smooth out high-frequency noise and maximize the signal of potential planets, which have PSFs that are roughly Gaussian\cite{Ruffio2017}. This cross correlation map is then used to estimate the noise level as a function of separation by computing the standard deviation of concentric annuli centered on the star. As we have a limited number of independent noise realizations, we correct for small sample statistics assuming the noise distribution is Gaussian \cite{Mawet2014} and compute the planet brightness that corresponds to the $5\sigma$-equivalent false positive probability of $2.9 \times 10^{-7}$, and we take this to be our sensitivity limit (i.e., achieved contrast). Thus, we produce a contrast curve for each planet spectral type that is uncalibrated for flux biases introduced by the data reduction steps described in this section and the previous section. 

For each type of planetary spectrum, we then perform two more stellar PSF subtractions using the same parameters as those used in Section \ref{sec:spec-psfsub}, but after injecting simulated planet signals. Simulated planets of that particular spectrum are injected as 2-D Gaussian signals and are used to quantify flux biases due to stellar PSF subtraction. For these simulated planet reductions, we inject planets at nine separations corresponding to the center of each of the nine annuli and at four different position angles so that the simulated planets spiral outward to avoid significantly influencing each other when using SDI. The second simulated planet reduction has the spiral arms offset by $45^{\circ}$ with respect to the first. We use the same Gaussian cross correlation routine that was used to estimate the noise level to measure the flux of each simulated planet after stellar PSF subtraction and calibrate flux measurement biases induced by our data reduction process. Then for each point in the contrast curve, we correct for these flux measurement biases. For the contrast within 40~pixels (566~mas), the calibration term is calculated as the linear interpolation between the two closest measured flux calibration factors. Outside of 40~pixels, the correction factor applied is the average correction factor for all simulated planets with separations greater than 40~pixels, as we determined empirically that the calibration factor is constant in this regime to measurement uncertainty ($\sim20\%$). This results in one T-type contrast curve and one L-type contrast curve for each dataset.

\subsection{Polarimetry Mode}

\subsubsection{Constructing Polarimetry Datacubes}
Instead of a 37-channel spectral datacube, we use the GPI DRP to generate datacubes where the third axis contains the two orthogonal polarizations of light. These polarimetry datacubes are an intermediate product that we then use for stellar PSF subtraction. The steps performed by the GPI DRP to generate polarimetry datacubes are listed in Table \ref{tab:podc}.

\begin{table}
\centering
\caption{GPI DRP processing steps for polarimetry datacubes.}
\begin{tabular}{| l | p{9cm} | }
 \hline
 \textbf{Primitive Name} & \textbf{Purpose}  \\
 \hline 
 Load Polarimetry Spot Calibration & Loads in the appropriate calibration file with the positions of the polarization spot pairs \\
 \hline
 Subtract Dark Background & Finds an appropriate dark frame and subtracts it from the raw 2-D image \\
 \hline
 Flexure 2D X Correlation with Polcal & Corrects for instrument flexure by cross correlating the spot calibration with the data and finding the optimal global offset in X and Y \\
 \hline
 Destripe Science Image & Models and removes vibration-induced microphonics\cite{Ingraham2014} \\
 \hline
 Interpolate Bad Pixels in 2D Frame & Identifies and fixes bad pixels\cite{Ingraham2014} \\
 \hline
 Assemble Polarization Cube & Constructs a 3-D datacube where the third dimension is the two orthogonal polarizations of light \\
 \hline
 Interpolate Bad Pixels in Cube & Fixes any remaining bad pixels using spatially nearby pixels \\
 \hline
  Correct Distortion & Corrects for optical distortion\cite{Konopacky2014} \\
 \hline
 Measure Star Position for Polarimetry & Locates the position of the star using a Radon-transform based algorithm\cite{Wang2014} \\
 \hline
 Measure Satellite Spot Flux in Polarimetry & Measures the flux of each satellite spot using aperture photometry. The aperture is elongated radially to match the shape of the satellite spots.\cite{Hung2016} \\
 \hline
 Measure Contrast in Pol Mode & Measures the contrast achieved in total intensity in a single frame of observation\cite{Millar-Blanchaer2016} \\
 \hline
\end{tabular}
\label{tab:podc}
\end{table}

\subsubsection{Stellar PSF Subtraction}
\label{sec:pol-psfsub}
For polarimetry data, we subtract the diffracted light of the host star using two methods: polarimetric differential imaging (PDI) to look for polarized scattered light from circumstellar dust around an unpolarized star and ADI to look for all scattered light from circumstellar material. 

To perform PDI, we use the GPI DRP to execute the steps listed in Table \ref{tab:pdi}. This outputs 3-D Stokes cubes where the third dimension contains the four components of the Stokes vectors. We also generate radial Stokes cubes by transforming the Stokes vectors into a radial Stokes basis\cite{Schmid2006}.  Since light from most debris disks is only scattered once, and thus has a tangential polarization, the radial Stokes basis typically has all of the polarized astrophysical signal in one Stokes parameter. For PDI, to account for instrumental polarization, we measure and subtract out the apparent stellar polarization by measuring the polarized signal behind the focal plane mask. One advantage of this technique is that it is robust against apparent stellar polarization due to interstellar polarization, unresolved disk structure, or other potential astrophysical polarization sources. Polarization from these sources would also affect the stellar speckles and needs to be subtracted out since we are only interested in the resolved polarized component in GPI's field of view. 

\begin{table}
\centering
\caption{GPI DRP processing steps for PDI.}
\begin{tabular}{| l | p{8cm} | }
 \hline
 \textbf{Primitive Name} & \textbf{Purpose}  \\
 \hline 
 Accumulate Images & Gathers together all the images in memory and demarcates that the following primitives will be run on a series of images, rather than on each image individually \\
 \hline
 Clean Polarization Pairs via Double Difference &  Removes polarized and unpolarized speckles in the images using data taken with different half-wave plate orientations\cite{Perrin2015} \\
 \hline
 Smooth a 3D Cube & Convolves each polarization image with a Gaussian with a FWHM of 1~pixel to suppress pixel-to-pixel noise and improve the noise properties in the final Stokes Cube without sacrificing significant spatial information. \\
 \hline
 Subtract Mean Stellar Polarization from podc & Uses an annulus inside the occulting mask between 7 and 13~pixels from the star to measure the stellar polarization and subtract it off\cite{Millar-Blanchaer2016}. The results are saved and can be referenced at a later time. \\
 \hline
 Rotate North Up & Rotates each image so that north is up and east is left \\
 \hline
 Combine Polarization Sequence & Takes the entire sequence and makes a single 3-D datacube where the third dimension is the four Stokes parameters. This Stokes cube is then saved to disk. \\
 \hline
 Convert Stokes Cube to Radial & Transforms the Stokes cube into a radial Stokes basis, which is also saved to disk \\
 \hline
 Get Statistics on Polarimetry Vectors & Generates a histogram of polarization directions across the entire field of view and saves the polarimetry quicklook display \\ 
 \hline
\end{tabular}
\label{tab:pdi}
\end{table}

To look for an unpolarized astrophysical signal next to the star in our polarimetry data, we sum the two orthogonal polarizations in each polarimetry cube and treat each cube as a single broadband image. Then, \texttt{pyKLIP} uses ADI to model and subtract out the stellar PSF using two different sets of reduction parameters. The images are not high-pass filtered before stellar PSF subtraction to preserve extended disk emission. The first reduction, the conservative one, divides the image into seven annuli, uses images where potential astrophysical sources have moved by at least 3~pixels due to ADI to be used to model the stellar PSF, and reconstructs the stellar PSF using 1, 3, 5, 10, 20, and 50 KL modes. The second reduction, the aggressive one, divides the image into nine annuli, uses images where potential astrophysical sources have moved by at least 1~pixel, and reconstructs the stellar PSF using 1, 2, 3, 4, 5, 10, 20, and 50 KL modes. We found that these two sets of parameters provide a conservative and aggressive reduction for studying circumstellar materials. The conservative reduction tries to mitigate stellar PSF subtraction biases on circumstellar material to preserve the diffuse emission and disk morphology. The aggressive reduction allows one to search for faint structure close in to the star.

\section{Data Cruncher Software Implementation}
\label{sec:code-data-cruncher}
Here, we will discuss how we implemented the Data Cruncher framework in software. In general, thread synchronization, network communication, and file system interrupts are key techniques for this implementation. 

\subsection{Processing Backend}
\label{sec:processing-backend}
The Processing Backend manages the data flow through the three data reduction pipelines discussed in Appendix \ref{sec:data-reduction} (GPI DRP, \texttt{pyKLIP}, and cADI). The Processing Backend is broken up into Python threads that each manage one task and are depicted as the boxes inside the Processing Backend box in Figure \ref{fig:data-cruncher-schematic}. The threads communicate with each other by passing messages through queues and use monitors (implemented using the \texttt{Lock()} and \texttt{Condition()} objects available in Python's \texttt{threading} library) as the synchronization construct to block threads that are waiting for new messages to be passed into the queue. For example, after one pipeline finishes processing some intermediate data products, it can send a message about these data products to another thread via their communication queue, alerting this second thread, which was previously sleeping since it had nothing to do, of the new data to process. The Processing Backend has four different kinds of threads: a network interface thread, the GPI DRP thread, stellar PSF subtraction threads, and the \texttt{GPIFileProcessor}, the backbone that manages the data flow. 

The network interface thread receives commands from Instructor processes. Instructors pass messages that contain the file or files to be processed, the output directory, whether the input data are raw 2-D images or 3-D datacubes, what kind of data it is (i.e., spectral, polarimetry, or calibration data), and optional parameters such as which stellar PSF subtraction algorithms to use. There are three network interface threads available: web socket, Message Passing Interface (MPI), or a regular queue. These threads are mutually exclusive so only one thread ever runs. The web socket thread is used in most contexts, except when run on a supercomputer. The network interface's main purpose is to parse the received messages and send the messages to the \texttt{GPIFileProcessor}.

The \texttt{GPIFileProcessor} thread manages all of the various pipeline threads. The \texttt{GPIFileProcessor} receives messages from the network interface and parses the instructions into work units that need to be passed into the various pipelines. It then checks against the database to ensure that the files are not marked as bad, discarding instructions for bad data as it goes along. All good, raw, 2-D data are sent to the GPI DRP for data processing. 

Because the GPI DRP was not designed to be fully automated to this degree and does not fully expose all of its features programmatically, a few workarounds were required. The GPI DRP uses a recipe directory, where each job is an XML file detailing the data reduction steps that need to be executed on some data. These recipe files exist in three states: waiting to be executed, in the processes of being executed, and finished being executed (either successful or failed).  The \texttt{GPIFileProcessor} writes XML recipe files into the queue directory, which queues a job for the GPI DRP. The recipe for science data are custom recipes specified in Appendix \ref{sec:data-reduction}. For calibration files, an instance of the GPI DRP Data Parser is created, and the Data Parser generates the appropriate recipes to be written into the queue.

Once a recipe file is written into the queue, we need to identify when the GPI DRP finished that recipe so we can then run the appropriate post-processing algorithms. The GPI DRP thread's sole purpose is to track this. When the GPI DRP finishes processing a recipe, it updates the recipe file to indicate whether the recipe succeeded or failed. Thus, the GPI DRP thread uses the Python \texttt{watchdog} package to receive file system interrupts. When a recipe file has been updated, the GPI DRP thread is awakened to run a function that notifies the \texttt{GPIFileProcessor} that data has been processed. 

Except in the case of quicklook reductions, the \texttt{GPIFileProcessor} waits on the condition that all raw data are finished being processed by the GPI DRP, regardless of whether all files were successfully processed, before running stellar PSF subtraction. When it is notified by the GPI DRP thread that all queued raw files have been processed, then it passes jobs to the stellar PSF subtraction threads. Generally, there are two stellar PSF subtraction threads that run in parallel: one for \texttt{pyKLIP} and one for cADI. Both threads run their respective pipelines as subprocesses rather than subthreads to avoid the Python global interpreter lock and to improve memory efficiency by relying on the operating system to free memory rather than the Python garbage collector. As cADI does not require many computational resources, it is practical to run the two threads in parallel. The cADI thread runs two reductions as discussed in Appendix \ref{sec:spec-psfsub}. The pyKLIP thread receives eight jobs for each spectral mode dataset: three different stellar PSF subtractions optimized for different science objectives as discussed in Appendix \ref{sec:spec-psfsub}, four reductions to inject and recover fake planets to calibrate out flux biases induced by the data processing, and one calculation of the contrast curve as detailed in Appendix \ref{sec:spec-contrastcurves}. These work units are broken up and prioritized so that if multiple datasets have stellar PSF subtractions queued up, a single KLIP reduction for each dataset is prioritized first, allowing for a fast initial look in real time during an observing night. For polarimetry mode data, two \texttt{pyKLIP} reductions (as discussed in Appendix \ref{sec:pol-psfsub}) are queued up to look for total intensity disk signal. PDI reductions are also queued into the GPI DRP queue to subtract the unpolarized stellar signal to look for polarized astrophysical emission. 

For quicklook reductions, only cADI is used for spectral mode data and the GPI DRP still performs the quicklook PDI reductions. Quicklook reductions are specified with a flag in the instructions received by the network interface. Upon seeing this flag, the \texttt{GPIFileProcessor} does not wait for all raw files to be finished by the GPI DRP. Rather, the goal is to have the quicklook reduction done as fast as possible in real time with however many files have already been processed. Because of this, when multiple files are downloaded by the Data Cruncher at once due to latency in the file syncing, duplicate quicklook stellar PSF subtraction jobs are created. To avoid having the same instruction multiple times in the stellar PSF subtraction queue to improve efficiency, the queue through which jobs are passed to the stellar PSF subtraction threads disallows duplicates. 

\subsection{Instructors}
\label{sec:instructors}
Currently, we use three different Instructor interfaces that send commands to the Processing Backends. Multiple Instructors can talk to the same Processing Backend, and to practically handle this, we do not leave web sockets open, closing them immediately after sending instructions so that another Instructor is able to connect without timing out. Often times, the Instructor and the Processing Backend live on the same machine as the Instructors do not consume much computing resources. 

The Realtime Scanner module handles all of the real-time processing. It uses the Python \texttt{watchdog} module to receive alerts from the operating system when new data are synced to the computer and written to disk. Upon being alerted of a new file, the Realtime Scanner decides how to process the file depending on the context in which the file is taken. To keep track of the state of observing, the logic inside of the Realtime Scanner is implemented as a finite state machine. The finite state machine logic handles almost all standard observing procedures for spectral, polarimetry, and calibration data taken as part of GPIES. The only exception are thermal background frames taken in $K$-band.  

The Reprocessor Instructor is a series of Python functions that can be called on demand to perform individual tasks such as reprocessing a single dataset on a target or processing a list of raw files. Each function uses the \texttt{mysql} Python library to query the database for the desired files requested by the user and sends commands to the Processing Backend to process the files appropriately. The Reprocessor can also query for all data from the campaign, generate instructions to reprocess all of the raw data from scratch, and save the commands to text files to be uploaded onto a supercomputing cluster that doesn't have direct database access.

The third Instructor is the controlling node for the Super Data Cruncher, which is what the Data Cruncher is called when it runs on a supercomputer. Each node on the supercomputer runs one instance of the Processing Backend. The previous two instructors use web sockets to communicate to the Processing Backend, but in supercomputer clusters, MPI is the network interface of choice. The exception is the controlling node itself, which uses a simple queue interface to pass instructions to its own Processing Backend, avoiding the overhead of MPI. The controlling node reads instructions that have been pre-generated in a text file to avoid setting up a connection to the database from the supercomputing center. It then distributes the instructions across all of the nodes with a granularity level of a full sequence on a single target.

Practically, the reprocessing of the entire campaign is done in two phases: once for the raw 2-D data to make datacubes, and once to run stellar PSF subtraction on the datacubes. As processing the raw data requires running the GPI DRP, and thus requiring IDL licenses for each node, we typically only use $\sim$20 nodes to do this step. Afterwards, a script is run to quality check the reductions, ensuring that calibration issues like flexure offsets are handled properly. Then, the Super Data Cruncher runs stellar PSF subtraction and contrast curve generation for all datasets. Due to IDL licensing issues, only a small subset of nodes are designated to run all of the cADI reductions, but since the cADI reductions are fast to compute, this does not slow down the reprocessing. \texttt{pyKLIP} stellar PSF subtraction and contrast curve generation is evenly distributed across all of the nodes.

\acknowledgments 
 
The Gemini Observatory is operated by the Association of Universities for Research in
Astronomy, Inc., under a cooperative agreement with the NSF on behalf of the Gemini
partnership: the National Science Foundation (United States), the National Research
Council (Canada), CONICYT (Chile), the Australian Research Council (Australia),
Minist\'erio da Ci\'encia, Tecnologia e Inova\c{c}\=ao (Brazil), and Ministerio de Ciencia,
Tecnolog\'ia e Innovaci\'on Productiva (Argentina). 
This work was supported in part by NASA's NExSS program, grant number NNX15AD95G.
This research used resources of the National Energy Research Scientific Computing Center, a DOE Office of Science User Facility supported by the Office of Science of the U.S. Department of Energy under Contract No. DE-AC02-05CH11231. 
This work used the Extreme Science and Engineering Discovery Environment (XSEDE), which is supported by National Science Foundation grant number ACI-1548562. Support for MMB's work was provided by NASA through Hubble Fellowship grant \#51378.01-A awarded by the Space Telescope Science Institute, which is operated by the Association of Universities for Research in Astronomy, Inc., for NASA, under contract NAS5-26555.
Portions of this work were performed under the auspices of the U.S. Department of Energy by Lawrence Livermore National Laboratory under Contract DE-AC52-07NA27344.

We thank the two anonymous referees for their thorough review and suggestions. We thank the NERSC and SDSC staff for their helpful support in providing computational resources and technical help. We also thank Barry Mieny for granting us permission to use the database icon used in Figure \ref{fig:data-infrastructure}. 
This research made use of Astropy, a community-developed core Python package for Astronomy\cite{Astropy}.

\bibliography{main} 

\begin{thebibliography}{10}

\bibitem{2007arXiv0704.1454G}
{Graham}, J.~R., {Macintosh}, B., {Doyon}, R., {Gavel}, D., {Larkin}, J.,
  {Levine}, M., {Oppenheimer}, B., {Palmer}, D., {Saddlemyer}, L.,
  {Sivaramakrishnan}, A., {Veran}, J.-P., and {Wallace}, K., ``{Ground-Based
  Direct Detection of Exoplanets with the Gemini Planet Imager (GPI)},'' {\em
  ArXiv e-prints}  (Apr. 2007).

\bibitem{Macintosh2014}
{Macintosh}, B., {Graham}, J.~R., {Ingraham}, P., {Konopacky}, Q., {Marois},
  C., {Perrin}, M., {Poyneer}, L., {Bauman}, B., {Barman}, T., {Burrows},
  A.~S., {Cardwell}, A., {Chilcote}, J., {De Rosa}, R.~J., {Dillon}, D.,
  {Doyon}, R., {Dunn}, J., {Erikson}, D., {Fitzgerald}, M.~P., {Gavel}, D.,
  {Goodsell}, S., {Hartung}, M., {Hibon}, P., {Kalas}, P., {Larkin}, J.,
  {Maire}, J., {Marchis}, F., {Marley}, M.~S., {McBride}, J.,
  {Millar-Blanchaer}, M., {Morzinski}, K., {Norton}, A., {Oppenheimer}, B.~R.,
  {Palmer}, D., {Patience}, J., {Pueyo}, L., {Rantakyro}, F., {Sadakuni}, N.,
  {Saddlemyer}, L., {Savransky}, D., {Serio}, A., {Soummer}, R.,
  {Sivaramakrishnan}, A., {Song}, I., {Thomas}, S., {Wallace}, J.~K.,
  {Wiktorowicz}, S., and {Wolff}, S., ``{First light of the Gemini Planet
  Imager},'' {\em Proceedings of the National Academy of Science}~{\bf 111},
  12661--12666 (Sept. 2014).

\bibitem{Poyneer2016}
{Poyneer}, L.~A., {Palmer}, D.~W., {Macintosh}, B., {Savransky}, D.,
  {Sadakuni}, N., {Thomas}, S., {V{\'e}ran}, J.-P., {Follette}, K.~B.,
  {Greenbaum}, A.~Z., {Mark Ammons}, S., {Bailey}, V.~P., {Bauman}, B.,
  {Cardwell}, A., {Dillon}, D., {Gavel}, D., {Hartung}, M., {Hibon}, P.,
  {Perrin}, M.~D., {Rantakyr{\"o}}, F.~T., {Sivaramakrishnan}, A., and {Wang},
  J.~J., ``{Performance of the Gemini Planet Imager's adaptive optics
  system},'' {\em \ao}~{\bf 55},  323 (Jan. 2016).

\bibitem{Soummer2011}
{Soummer}, R., {Sivaramakrishnan}, A., {Pueyo}, L., {Macintosh}, B., and
  {Oppenheimer}, B.~R., ``{Apodized Pupil Lyot Coronagraphs for Arbitrary
  Apertures. III. Quasi-achromatic Solutions},'' {\em \apj}~{\bf 729},  144
  (Mar. 2011).

\bibitem{Chilcote2012}
{Chilcote}, J.~K., {Larkin}, J.~E., {Maire}, J., {Perrin}, M.~D., {Fitzgerald},
  M.~P., {Doyon}, R., {Thibault}, S., {Bauman}, B., {Macintosh}, B.~A.,
  {Graham}, J.~R., and {Saddlemyer}, L., ``{Performance of the integral field
  spectrograph for the Gemini Planet Imager},'' in [{\em Ground-based and
  Airborne Instrumentation for Astronomy IV}{\nolinebreak\hspace{0.1em}]},
  {\em \procspie} {\bf 8446},  84468W (Sept. 2012).

\bibitem{Larkin2014}
{Larkin}, J.~E., {Chilcote}, J.~K., {Aliado}, T., {Bauman}, B.~J., {Brims}, G.,
  {Canfield}, J.~M., {Cardwell}, A., {Dillon}, D., {Doyon}, R., {Dunn}, J.,
  {Fitzgerald}, M.~P., {Graham}, J.~R., {Goodsell}, S., {Hartung}, M., {Hibon},
  P., {Ingraham}, P., {Johnson}, C.~A., {Kress}, E., {Konopacky}, Q.~M.,
  {Macintosh}, B.~A., {Magnone}, K.~G., {Maire}, J., {McLean}, I.~S., {Palmer},
  D., {Perrin}, M.~D., {Quiroz}, C., {Rantakyr{\"o}}, F., {Sadakuni}, N.,
  {Saddlemyer}, L., {Serio}, A., {Thibault}, S., {Thomas}, S.~J., {Vallee}, P.,
  and {Weiss}, J.~L., ``{The integral field spectrograph for the Gemini planet
  imager},'' in [{\em Ground-based and Airborne Instrumentation for Astronomy
  V}{\nolinebreak\hspace{0.1em}]},  {\em \procspie} {\bf 9147},  91471K (July
  2014).

\bibitem{Perrin2015}
{Perrin}, M.~D., {Duchene}, G., {Millar-Blanchaer}, M., {Fitzgerald}, M.~P.,
  {Graham}, J.~R., {Wiktorowicz}, S.~J., {Kalas}, P.~G., {Macintosh}, B.,
  {Bauman}, B., {Cardwell}, A., {Chilcote}, J., {De Rosa}, R.~J., {Dillon}, D.,
  {Doyon}, R., {Dunn}, J., {Erikson}, D., {Gavel}, D., {Goodsell}, S.,
  {Hartung}, M., {Hibon}, P., {Ingraham}, P., {Kerley}, D., {Konapacky}, Q.,
  {Larkin}, J.~E., {Maire}, J., {Marchis}, F., {Marois}, C., {Mittal}, T.,
  {Morzinski}, K.~M., {Oppenheimer}, B.~R., {Palmer}, D.~W., {Patience}, J.,
  {Poyneer}, L., {Pueyo}, L., {Rantakyr{\"o}}, F.~T., {Sadakuni}, N.,
  {Saddlemyer}, L., {Savransky}, D., {Soummer}, R., {Sivaramakrishnan}, A.,
  {Song}, I., {Thomas}, S., {Wallace}, J.~K., {Wang}, J.~J., and {Wolff},
  S.~G., ``{Polarimetry with the Gemini Planet Imager: Methods, Performance at
  First Light, and the Circumstellar Ring around HR 4796A},'' {\em \apj}~{\bf
  799},  182 (Feb. 2015).

\bibitem{Marois2006_adi}
{Marois}, C., {Lafreni{\`e}re}, D., {Doyon}, R., {Macintosh}, B., and {Nadeau},
  D., ``{Angular Differential Imaging: A Powerful High-Contrast Imaging
  Technique},'' {\em \apj}~{\bf 641},  556--564 (Apr. 2006).

\bibitem{Marois2000}
{Marois}, C., {Doyon}, R., {Racine}, R., and {Nadeau}, D., ``{Efficient Speckle
  Noise Attenuation in Faint Companion Imaging},'' {\em \pasp}~{\bf 112},
  91--96 (Jan. 2000).

\bibitem{Marois2014}
{Marois}, C., {Correia}, C., {Galicher}, R., {Ingraham}, P., {Macintosh}, B.,
  {Currie}, T., and {De Rosa}, R., ``{GPI PSF subtraction with TLOCI: the next
  evolution in exoplanet/disk high-contrast imaging},'' in [{\em Adaptive
  Optics Systems IV}{\nolinebreak\hspace{0.1em}]},  {\em \procspie} {\bf 9148},
   91480U (July 2014).

\bibitem{xsede}
Towns, J., Cockerill, T., Dahan, M., Foster, I., Gaither, K., Grimshaw, A.,
  Hazlewood, V., Lathrop, S., Lifka, D., Peterson, G.~D., Roskies, R., Scott,
  J.~R., and Wilkins-Diehr, N., ``Xsede: Accelerating scientific discovery,''
  {\em Computing in Science \& Engineering}~{\bf 16}(5),  62--74 (2014).

\bibitem{mcbride2011experimental}
McBride, J., Graham, J., Macintosh, B., Beckwith, S., Marois, C., Poyneer, L.,
  and Wiktorowicz, S., ``{Experimental Design for the Gemini Planet Imager},''
  {\em Publications of the Astronomical Society of the Pacific}~{\bf 123},
  692--708 (2011).

\bibitem{savransky2013campaign}
Savransky, D., Macintosh, B.~A., Graham, J., and Konopacky, Q.~M., ``Campaign
  scheduling and analysis for the {Gemini Planet Imager},'' in [{\em
  Proceedings of the International Astronomical
  Union}{\nolinebreak\hspace{0.1em}]},   {\bf 8}(S299),  68--69, Cambridge Univ
  Press (2013).

\bibitem{Bailey2016}
{Bailey}, V.~P., {Poyneer}, L.~A., {Macintosh}, B.~A., {Savransky}, D., {Wang},
  J.~J., {De Rosa}, R.~J., {Follette}, K.~B., {Ammons}, S.~M., {Hayward}, T.,
  {Ingraham}, P., {Maire}, J., {Palmer}, D.~W., {Perrin}, M.~D., {Rajan}, A.,
  {Rantakyr{\"o}}, F.~T., {Thomas}, S., and {V{\'e}ran}, J.-P., ``{Status and
  performance of the Gemini Planet Imager adaptive optics system},'' in [{\em
  Adaptive Optics Systems V}{\nolinebreak\hspace{0.1em}]},  {\em \procspie}
  {\bf 9909},  99090V (July 2016).

\bibitem{Konopacky2016}
{Konopacky}, Q.~M., {Rameau}, J., {Duch{\^e}ne}, G., {Filippazzo}, J.~C.,
  {Giorla Godfrey}, P.~A., {Marois}, C., {Nielsen}, E.~L., {Pueyo}, L.,
  {Rafikov}, R.~R., {Rice}, E.~L., {Wang}, J.~J., {Ammons}, S.~M., {Bailey},
  V.~P., {Barman}, T.~S., {Bulger}, J., {Bruzzone}, S., {Chilcote}, J.~K.,
  {Cotten}, T., {Dawson}, R.~I., {De Rosa}, R.~J., {Doyon}, R., {Esposito},
  T.~M., {Fitzgerald}, M.~P., {Follette}, K.~B., {Goodsell}, S., {Graham},
  J.~R., {Greenbaum}, A.~Z., {Hibon}, P., {Hung}, L.-W., {Ingraham}, P.,
  {Kalas}, P., {Lafreni{\`e}re}, D., {Larkin}, J.~E., {Macintosh}, B.~A.,
  {Maire}, J., {Marchis}, F., {Marley}, M.~S., {Matthews}, B.~C., {Metchev},
  S., {Millar-Blanchaer}, M.~A., {Oppenheimer}, R., {Palmer}, D.~W.,
  {Patience}, J., {Perrin}, M.~D., {Poyneer}, L.~A., {Rajan}, A.,
  {Rantakyr{\"o}}, F.~T., {Savransky}, D., {Schneider}, A.~C.,
  {Sivaramakrishnan}, A., {Song}, I., {Soummer}, R., {Thomas}, S., {Wallace},
  J.~K., {Ward-Duong}, K., {Wiktorowicz}, S.~J., and {Wolff}, S.~G.,
  ``{Discovery of a Substellar Companion to the Nearby Debris Disk Host HR
  2562},'' {\em \apjl}~{\bf 829},  L4 (Sept. 2016).

\bibitem{Millar-Blanchaer2017}
Millar-Blanchaer, M.~A., Esposito, T.~M., Fitzgerald, M.~P., Kalas, P., Perrin,
  M.~D., Macintosh, B., Graham, J.~R., and the GPIES~Collaboration, ``High
  contrast observations of circumstellar disks with the gemini planet imager's
  polariemtry mode,'' in [{\em Polarization Science and Remote Sensing
  VIII}{\nolinebreak\hspace{0.1em}]},  {\em \procspie} {\bf 10407} (2017).

\bibitem{Ruffio2017}
{Ruffio}, J.-B., {Macintosh}, B., {Wang}, J.~J., {Pueyo}, L., {Nielsen}, E.~L.,
  {De Rosa}, R.~J., {Czekala}, I., {Marley}, M.~S., {Arriaga}, P., {Bailey},
  V.~P., {Barman}, T., {Bulger}, J., {Chilcote}, J., {Cotten}, T., {Doyon}, R.,
  {Duch{\^e}ne}, G., {Fitzgerald}, M.~P., {Follette}, K.~B., {Gerard}, B.~L.,
  {Goodsell}, S.~J., {Graham}, J.~R., {Greenbaum}, A.~Z., {Hibon}, P., {Hung},
  L.-W., {Ingraham}, P., {Kalas}, P., {Konopacky}, Q., {Larkin}, J.~E.,
  {Maire}, J., {Marchis}, F., {Marois}, C., {Metchev}, S., {Millar-Blanchaer},
  M.~A., {Morzinski}, K.~M., {Oppenheimer}, R., {Palmer}, D., {Patience}, J.,
  {Perrin}, M., {Poyneer}, L., {Rajan}, A., {Rameau}, J., {Rantakyr{\"o}},
  F.~T., {Savransky}, D., {Schneider}, A.~C., {Sivaramakrishnan}, A., {Song},
  I., {Soummer}, R., {Thomas}, S., {Wallace}, J.~K., {Ward-Duong}, K.,
  {Wiktorowicz}, S., and {Wolff}, S., ``{Improving and Assessing Planet
  Sensitivity of the GPI Exoplanet Survey with a Forward Model Matched
  Filter},'' {\em \apj}~{\bf 842},  14 (June 2017).

\bibitem{Perrin2014}
{Perrin}, M.~D., {Maire}, J., {Ingraham}, P., {Savransky}, D.,
  {Millar-Blanchaer}, M., {Wolff}, S.~G., {Ruffio}, J.-B., {Wang}, J.~J.,
  {Draper}, Z.~H., {Sadakuni}, N., {Marois}, C., {Rajan}, A., {Fitzgerald},
  M.~P., {Macintosh}, B., {Graham}, J.~R., {Doyon}, R., {Larkin}, J.~E.,
  {Chilcote}, J.~K., {Goodsell}, S.~J., {Palmer}, D.~W., {Labrie}, K.,
  {Beaulieu}, M., {De Rosa}, R.~J., {Greenbaum}, A.~Z., {Hartung}, M., {Hibon},
  P., {Konopacky}, Q., {Lafreniere}, D., {Lavigne}, J.-F., {Marchis}, F.,
  {Patience}, J., {Pueyo}, L., {Rantakyr{\"o}}, F.~T., {Soummer}, R.,
  {Sivaramakrishnan}, A., {Thomas}, S., {Ward-Duong}, K., and {Wiktorowicz},
  S., ``{Gemini Planet Imager observational calibrations I: Overview of the GPI
  data reduction pipeline},'' in [{\em Ground-based and Airborne
  Instrumentation for Astronomy V}{\nolinebreak\hspace{0.1em}]},  {\em
  \procspie} {\bf 9147},  91473J (July 2014).

\bibitem{Perrin2016}
{Perrin}, M.~D., {Ingraham}, P., {Follette}, K.~B., {Maire}, J., {Wang}, J.~J.,
  {Savransky}, D., {Arriaga}, P., {Bailey}, V.~P., {Bruzzone}, S., {Chilcote},
  J.~K., {De Rosa}, R.~J., {Draper}, Z.~H., {Fitzgerald}, M.~P., {Greenbaum},
  A.~Z., {Hung}, L.-W., {Konopacky}, Q., {Macintosh}, B., {Marchis}, F.,
  {Marois}, C., {Millar-Blanchaer}, M.~A., {Nielsen}, E., {Rajan}, A.,
  {Rameau}, J., {Rantakyro}, F.~T., {Ruffio}, J.-B., {Ward-Duong}, K., {Wolff},
  S.~G., and {Zalesky}, J., ``{Gemini Planet Imager observational calibrations
  XI: pipeline improvements and enhanced calibrations after two years on
  sky},'' in [{\em Ground-based and Airborne Instrumentation for Astronomy
  VI}{\nolinebreak\hspace{0.1em}]},  {\em \procspie} {\bf 9908},  990837 (Aug.
  2016).

\bibitem{Ingraham2014}
{Ingraham}, P., {Perrin}, M.~D., {Sadakuni}, N., {Ruffio}, J.-B., {Maire}, J.,
  {Chilcote}, J., {Larkin}, J., {Marchis}, F., {Galicher}, R., and {Weiss}, J.,
  ``{Gemini planet imager observational calibrations II: detector performance
  and calibration},'' in [{\em Ground-based and Airborne Instrumentation for
  Astronomy V}{\nolinebreak\hspace{0.1em}]},  {\em \procspie} {\bf 9147},
  91477O (July 2014).

\bibitem{Wolff2014}
{Wolff}, S.~G., {Perrin}, M.~D., {Maire}, J., {Ingraham}, P.~J.,
  {Rantakyr{\"o}}, F.~T., and {Hibon}, P., ``{Gemini planet imager
  observational calibrations IV: wavelength calibration and flexure correction
  for the integral field spectograph},'' in [{\em Ground-based and Airborne
  Instrumentation for Astronomy V}{\nolinebreak\hspace{0.1em}]},  {\em
  \procspie} {\bf 9147},  91477H (Aug. 2014).

\bibitem{Marois2006_satspots}
{Marois}, C., {Lafreni{\`e}re}, D., {Macintosh}, B., and {Doyon}, R.,
  ``{Accurate Astrometry and Photometry of Saturated and Coronagraphic Point
  Spread Functions},'' {\em \apj}~{\bf 647},  612--619 (Aug. 2006).

\bibitem{Sivaramakrishnan2006}
{Sivaramakrishnan}, A. and {Oppenheimer}, B.~R., ``{Astrometry and Photometry
  with Coronagraphs},'' {\em \apj}~{\bf 647},  620--629 (Aug. 2006).

\bibitem{savransky2013}
Savransky, D., Thomas, S.~J., Poyneer, L.~A., and Macintosh, B.~A., ``Computer
  vision applications for coronagraphic optical alignment and image
  processing,'' {\em Applied Optics}~{\bf 52}(14),  3394--3403 (2013).

\bibitem{Wang2014}
{Wang}, J.~J., {Rajan}, A., {Graham}, J.~R., {Savransky}, D., {Ingraham},
  P.~J., {Ward-Duong}, K., {Patience}, J., {De Rosa}, R.~J., {Bulger}, J.,
  {Sivaramakrishnan}, A., {Perrin}, M.~D., {Thomas}, S.~J., {Sadakuni}, N.,
  {Greenbaum}, A.~Z., {Pueyo}, L., {Marois}, C., {Oppenheimer}, B.~R., {Kalas},
  P., {Cardwell}, A., {Goodsell}, S., {Hibon}, P., and {Rantakyr{\"o}}, F.~T.,
  ``{Gemini planet imager observational calibrations VIII: characterization and
  role of satellite spots},'' in [{\em Ground-based and Airborne
  Instrumentation for Astronomy V}{\nolinebreak\hspace{0.1em}]},  {\em
  \procspie} {\bf 9147},  914755 (July 2014).

\bibitem{Wang2015}
{Wang}, J.~J., {Ruffio}, J.-B., {De Rosa}, R.~J., {Aguilar}, J., {Wolff},
  S.~G., and {Pueyo}, L., ``{pyKLIP: PSF Subtraction for Exoplanets and
  Disks}.'' Astrophysics Source Code Library (June 2015).

\bibitem{Soummer2012}
{Soummer}, R., {Pueyo}, L., and {Larkin}, J., ``{Detection and Characterization
  of Exoplanets and Disks Using Projections on Karhunen-Lo{\`e}ve
  Eigenimages},'' {\em \apjl}~{\bf 755},  L28 (Aug. 2012).

\bibitem{Pueyo2015}
{Pueyo}, L., {Soummer}, R., {Hoffmann}, J., {Oppenheimer}, R., {Graham}, J.~R.,
  {Zimmerman}, N., {Zhai}, C., {Wallace}, J.~K., {Vescelus}, F., {Veicht}, A.,
  {Vasisht}, G., {Truong}, T., {Sivaramakrishnan}, A., {Shao}, M., {Roberts},
  Jr., L.~C., {Roberts}, J.~E., {Rice}, E., {Parry}, I.~R., {Nilsson}, R.,
  {Lockhart}, T., {Ligon}, E.~R., {King}, D., {Hinkley}, S., {Hillenbrand}, L.,
  {Hale}, D., {Dekany}, R., {Crepp}, J.~R., {Cady}, E., {Burruss}, R.,
  {Brenner}, D., {Beichman}, C., and {Baranec}, C., ``{Reconnaissance of the HR
  8799 Exosolar System. II. Astrometry and Orbital Motion},'' {\em \apj}~{\bf
  803},  31 (Apr. 2015).

\bibitem{Mawet2014}
{Mawet}, D., {Milli}, J., {Wahhaj}, Z., {Pelat}, D., {Absil}, O., {Delacroix},
  C., {Boccaletti}, A., {Kasper}, M., {Kenworthy}, M., {Marois}, C.,
  {Mennesson}, B., and {Pueyo}, L., ``{Fundamental Limitations of High Contrast
  Imaging Set by Small Sample Statistics},'' {\em \apj}~{\bf 792},  97 (Sept.
  2014).

\bibitem{Konopacky2014}
{Konopacky}, Q.~M., {Thomas}, S.~J., {Macintosh}, B.~A., {Dillon}, D.,
  {Sadakuni}, N., {Maire}, J., {Fitzgerald}, M., {Hinkley}, S., {Kalas}, P.,
  {Esposito}, T., {Marois}, C., {Ingraham}, P.~J., {Marchis}, F., {Perrin},
  M.~D., {Graham}, J.~R., {Wang}, J.~J., {De Rosa}, R.~J., {Morzinski}, K.,
  {Pueyo}, L., {Chilcote}, J.~K., {Larkin}, J.~E., {Fabrycky}, D., {Goodsell},
  S.~J., {Oppenheimer}, B.~R., {Patience}, J., {Saddlemyer}, L., and
  {Sivaramakrishnan}, A., ``{Gemini planet imager observational calibrations V:
  astrometry and distortion},'' in [{\em Ground-based and Airborne
  Instrumentation for Astronomy V}{\nolinebreak\hspace{0.1em}]},  {\em
  \procspie} {\bf 9147},  914784 (July 2014).

\bibitem{Hung2016}
{Hung}, L.-W., {Bruzzone}, S., {Millar-Blanchaer}, M.~A., {Wang}, J.~J.,
  {Arriaga}, P., {Metchev}, S., {Fitzgerald}, M.~P., {Sivaramakrishnan}, A.,
  and {Perrin}, M.~D., ``{Gemini planet imager observational calibration XII:
  photometric calibration in the polarimetry mode},'' in [{\em Ground-based and
  Airborne Instrumentation for Astronomy VI}{\nolinebreak\hspace{0.1em}]},
  {\em \procspie} {\bf 9908},  99083A (Aug. 2016).

\bibitem{Millar-Blanchaer2016}
{Millar-Blanchaer}, M.~A., {Perrin}, M.~D., {Hung}, L.-W., {Fitzgerald}, M.~P.,
  {Wang}, J.~J., {Chilcote}, J., {Graham}, J.~R., {Bruzzone}, S., and {Kalas},
  P.~G., ``{GPI observational calibrations XIV: polarimetric contrasts and new
  data reduction techniques},'' in [{\em Ground-based and Airborne
  Instrumentation for Astronomy VI}{\nolinebreak\hspace{0.1em}]},   {\bf 9908},
   990836 (Aug. 2016).

\bibitem{Schmid2006}
{Schmid}, H.~M., {Joos}, F., and {Tschan}, D., ``{Limb polarization of Uranus
  and Neptune. I. Imaging polarimetry and comparison with analytic models},''
  {\em \aap}~{\bf 452},  657--668 (June 2006).

\bibitem{Astropy}
{Astropy Collaboration}, {Robitaille}, T.~P., {Tollerud}, E.~J., {Greenfield},
  P., {Droettboom}, M., {Bray}, E., {Aldcroft}, T., {Davis}, M., {Ginsburg},
  A., {Price-Whelan}, A.~M., {Kerzendorf}, W.~E., {Conley}, A., {Crighton}, N.,
  {Barbary}, K., {Muna}, D., {Ferguson}, H., {Grollier}, F., {Parikh}, M.~M.,
  {Nair}, P.~H., {Unther}, H.~M., {Deil}, C., {Woillez}, J., {Conseil}, S.,
  {Kramer}, R., {Turner}, J.~E.~H., {Singer}, L., {Fox}, R., {Weaver}, B.~A.,
  {Zabalza}, V., {Edwards}, Z.~I., {Azalee Bostroem}, K., {Burke}, D.~J.,
  {Casey}, A.~R., {Crawford}, S.~M., {Dencheva}, N., {Ely}, J., {Jenness}, T.,
  {Labrie}, K., {Lim}, P.~L., {Pierfederici}, F., {Pontzen}, A., {Ptak}, A.,
  {Refsdal}, B., {Servillat}, M., and {Streicher}, O., ``{Astropy: A community
  Python package for astronomy},'' {\em \aap}~{\bf 558},  A33 (Oct. 2013).

\end{thebibliography}
\bibliographystyle{spiebib} 

\vspace{2ex}\noindent\textbf{Jason Wang} is a PhD student in the Astronomy Department at the University of California, Berkeley. His current research interests include characterizing directly-imaged exoplanets, image post-processing techniques, and software development for astronomy.

\vspace{1ex}
\noindent Biographies and photographs of the other authors are not available.

\listoffigures
\listoftables

\end{spacing}
\end{document}